\documentclass[aps,prd,reprint,twocolumn,superscriptaddress,showpacs,nofootinbib,notitlepage]{revtex4-1}
\usepackage[utf8]{inputenc}
\usepackage{graphicx}
\usepackage{latexsym,amsmath,amssymb,lmodern,float,url}
\usepackage{natbib}
\usepackage{color}
\usepackage{microtype}
\usepackage{slashed}
\usepackage{multirow}
\usepackage{comment}
\usepackage{bbm}
\usepackage{tikz}
\usepackage{qcircuit}

\usepackage{etoolbox}
\apptocmd{\sloppy}{\hbadness 10000\relax}{}{}

\newcommand{\eq}[1]{Eq.~(\ref{eq:#1})}
\newcommand{\fig}[1]{Fig.~(\ref{fig:#1})}

\newcommand{\id}{\mathbbm 1}

\usepackage[colorlinks=true,backref=false, linktocpage=true,
citecolor=blue,urlcolor=blue,linkcolor=blue,pdfpagemode=UseOutlines]{hyperref}

\hypersetup{%
  bookmarksnumbered=true,
  pdftitle = {},
  pdfsubject = {},
  pdfauthor = {},
  pdfkeywords = {}
}

\let\Re\undefined
\let\Im\undefined
\DeclareMathOperator{\Tr}{Tr}
\DeclareMathOperator{\Re}{Re}
\DeclareMathOperator{\Im}{Im}

\renewcommand{\d}{\mathrm d}

\begin{document}
\title{General Methods for Digital Quantum Simulation of Gauge Theories}
\affiliation{Department of Physics, University of Maryland, College Park, MD 20742, USA}
\author{Henry Lamm}
\email{hlamm@umd.edu}
\affiliation{Department of Physics, University of Maryland, College Park, MD 20742, USA}
\author{Scott Lawrence}
\email{srl@umd.edu}
\affiliation{Department of Physics, University of Maryland, College Park, MD 20742, USA}
\author{Yukari Yamauchi}
\email{yyukari@umd.edu}
\affiliation{Department of Physics, University of Maryland, College Park, MD 20742, USA}
\date{\today}

\collaboration{NuQS Collaboration}
\begin{abstract}
A general scheme is presented for simulating gauge theories, with
matter fields, on a digital quantum computer. A Trotterized time-evolution operator that respects gauge symmetry is constructed, and a procedure for obtaining time-separated, gauge-invariant correlators is detailed.
We demonstrate the procedure on small lattices, including the simulation of a
2+1D non-Abelian gauge theory.
\end{abstract}

\maketitle

\section{Introduction}
Quantum simulations are motivated by inherent obstacles to the classical, nonperturbative simulation of quantum field theories~\cite{Feynman:1981tf}.  Deterministic methods struggle with exponential state spaces. Sign problems stymie Monte Carlo methods both at finite fermion density~\cite{Troyer:2004ge,Alexandru:2018ddf} and in real-time evolution~\cite{Alexandru:2016gsd}.  While large-scale quantum computers will greatly enhance calculations in quantum field theory, for the foreseeable future quantum computers are limited to tens or hundreds of non-error-corrected qubits with circuit depths fewer than a 1000 gates --- the so-called Noisy Intermediate-Scale Quantum (NISQ) era.  Along with hardware development, theoretical issues impede full use of quantum computers.  Despite these limits, algorithms have been demonstrated in toy field theories~\cite{Dumitrescu:2018njn,Roggero:2018hrn,Lu:2018pjk,Martinez:2016yna,Kokail:2018eiw,Klco:2018kyo,Lamm:2018siq,Macridin:2018gdw}.  Viable quantum simulations require addressing four interconnected issues: representation, preparation, evaluation, and propagation.  Proposals in the literature address one or more of these topics.  For gauge theories, additional complications arise which we discuss here.

The first hurdle is the representation of a gauge field in a set of qubits, or quantum register.  While natural matchings exist for fermionic fields~\cite{Jordan:1928wi,Verstraete:2005pn,Zohar:2018cwb,2016PhRvA..94c0301W}, the digitization of a bosonic field is nontrivial.  This is reminiscent of the early days of classical lattice field theory where memory resources limited calculations.  Ideas include approximating by: finite subsets~\cite{Hackett:2018cel,Alexandru:2019nsa}, Fock-state truncation~\cite{Macridin:2018gdw,Yeter-Aydeniz:2018mix,Klco:2018zqz}, dual variables~\cite{Bazavov:2015kka,Zhang:2018ufj,Unmuth-Yockey:2018xak,Unmuth-Yockey:2018ugm,Kaplan:2018vnj,Meurice:2019ddf}, optical representations~\cite{Zache:2018jbt}, and the prepotential formulation~\cite{Raychowdhury:2018osk}.  A common suggestion is to limit the register to physical states by gauge fixing, at the cost of increased circuit depth in the time-evolution, although a practical method for nonabelian theories remains undescribed.

Given a register, the next concern is the preparation of desired quantum states.  In strongly-coupled theories, how the asymptotic states depend upon the fundamental fields is unknown.  Most methods focus on ground state construction.  Examples include using quantum variational methods~\cite{peruzzo2014variational,Kokail:2018eiw}, Quantum Phase Estimation (QPE)~\cite{Abrams:1998pd,nielsen2000quantum,PhysRevLett.117.010503}, Quantum Adiabatic Algorithm (QAA)~\cite{farhi2000quantum,Farhi472}, and spectral combing~\cite{Kaplan:2017ccd}.  Some preparation methods require efficient real-time evolution. Classical-quantum hybrid methods~\cite{Lamm:2018siq} and dimensional reduction~\cite{2010PhRvL.105q0405B} have been proposed to initializing thermal states.  Simulations of scattering are hindered by state preparation~\cite{Jordan:2011ne,Jordan:2011ci,Garcia-Alvarez:2014uda,Jordan:2014tma,Jordan:2017lea,Moosavian:2017tkv}. 

Evaluation of observables represented by a single, hermitian operator at a single time is straight forward.  This includes most static properties.  Other observables, like time-separated composite operators (e.g. parton distribution functions) are more complicated.  Naively, the first nonunitary operator collapses the quantum state.  One resolution introduces ancillary probe and control qubits~\cite{PhysRevLett.113.020505}, which we use here.  QPE~\cite{Abrams:1998pd} has been used to compute linear response~\cite{Roggero:2018hrn}.  Another intriguing idea uses quantum sensors to implement generating functionals~\cite{Bermudez:2017yrq}.   A second problem arises from state contamination.  In Euclidean lattice field theory, we can use imaginary-time evolution to separate states overlapping with the same operator.  Real-time evolution lacks this separation, so we need novel methods to project onto desired states, perhaps through a distillation-like technique~\cite{Peardon:2009gh}.

In this work, we construct a method for mapping the Hilbert space of a gauge theory onto the Hilbert space of a quantum computer, and simulating its propagation. Propagating a system for a time $t$ requires a unitary operator $\mathcal U(t)=e^{-iHt}$, which cannot be efficiently constructed on a quantum computer.  Instead, one Trotterizes the evolution by $\mathcal U(t)\approx (e^{-iH \frac{t}{N}})^N$ which allows efficient simulation~\cite{Jordan:2011ne,Jordan:2011ci,Jordan:2017lea,Garcia-Alvarez:2014uda,Jordan:2014tma,Moosavian:2017tkv,haah2018quantum}.  Trotterized versions of $\mathcal{U}(t)$ exist for electron-phonon systems~\cite{Macridin:2018gdw}, the abelian theories~\cite{Banerjee:2012pg,Hauke:2013jga,Kuno:2014npa,Kuno:2016ipi,Martinez:2016yna,Marcos:2014lda,Klco:2018kyo,Zache:2018jbt,Muschik:2016tws,Bazavov:2015kka,Zhang:2018ufj,Unmuth-Yockey:2018xak,Unmuth-Yockey:2018ugm,Gustafson:2019mpk,Zohar:2012ay}, and nonabelian theories~\cite{Byrnes:2005qx,Banerjee:2012xg,Wiese:2013uua,Zohar:2012xf,Tagliacozzo:2012df,Zohar:2015hwa,Zohar:2016iic,Bender:2018rdp}.  Each depends on the register used.  

Violations of gauge invariance in the time-evolution are potentially introduced through noisy gates, digitization, and Trotterization.  To remedy these, \cite{Stryker:2018efp} proposed using oracles to project into the physical subspace.  In this work, we construct a Trotterized time-evolution without gauge-fixing, for arbitrary gauge theories with coupled matter fields. Gauge invariance is exactly preserved after Trotterization. We connect this procedure to the transfer matrix formalism~\cite{Luscher:1976ms,Mitrjushkin:2002nk,Creutz:1999zy,Caracciolo:2012qw}.  The effect of noisy gates on gauge-invariance are hardware-dependent, and we leave this for future work. The error introduced by field digitization can be treated without impacting the algorithm.

A final, overarching concern is renormalization.  Work in quantum computing~\cite{Haegeman:2011uy,Verstraete:2010ft} and elsewhere~\cite{Rossi:2018zkn,Briceno:2017cpo} suggests that the renormalization factors may be cheaply computed classically, allowing for the matching of the bare parameters on the quantum computer to renormalized and physical ones.  In the context of the method considered here, a mapping between the Trotterization and a Euclidean action may allow renormalized couplings to be computed efficiently on a classical computer with standard lattice methods.  The transfer matrix formalism naturally provides such a mapping.  Other quantum-computing based proposals exist as well~\cite{Cotler:2018ehb,Cotler:2018ufx}.

This paper is structured as follows: in Sec.~\ref{sec:prerequisites} we describe the prerequisite registers and fundamental gates for our construction.  With these, Sec.~\ref{sec:yang-mills} formulates gauge-invariant $\mathcal{U}(t)$ for the gauge fields.  Sec.~\ref{sec:path-integral} links this procedure to the transfer matrix formalism, and outlines how renormalization may be performed classically. Sec.~\ref{sec:correlators} explains how nontrivial correlators may be computed.  Sec.~\ref{sec:scalar} and \ref{sec:spinor} extend the formulation to include scalar and spinor fields respectively.  Sec.~\ref{sec:demonstration} demonstrates of our method in the non-Abelian $D_4$ theory and the $\mathbb Z_2$ theory with staggered fermions.  We conclude in Sec.~\ref{sec:discussion}. 
\section{Prerequisites}\label{sec:prerequisites}
Developers of quantum algorithms assume idealized qubits, and certain unitary gates operating on those qubits (a sufficient universal gateset, for example, is the Hadamard gate, the $\frac\pi 8$ gate ($T$), and the controlled-not gate \texttt{CNOT}).  This follows a pattern in classical computing, where certain constructs (e.g., floating-point numbers and arrays) are assumed to specify an algorithm.  The implication of this is that \emph{any} correct implementation of the primitive constructs will do: the correctness of the quantum algorithm does not depend on whether superconducting qubits or trapped ions are used.

In this section, we describe a set of higher-level constructs that allows the simulation of a general nonabelian gauge theory. We outline how they might be implemented in practice, from the lower-level qubit/gateset primitives. Hopefully these specifications are helpful to focus the implementation of those tools that are most critical.

We will simulate a theory with a gauge group $G$. The central construct is the $G$-register, analogous to a classical variable storing an element of $G$. The Hilbert space of an ideal $G$-register is $\mathcal H_G = \mathbb C G$, the complex vector space with one basis element $\left|g\right>$ for every element $g \in G$. (Equivalently, it is the space of square-integrable functions from $G$ to $\mathbb C$.) The Hilbert space of multiple $G$-registers is constructed as a tensor product, so a computer with $N$ $G$-registers has Hilbert space $\mathcal H_G^{\otimes N}$.

For discrete groups, ideal $G$-registers are easy to implement by virtue of their finite number of elements; for instance, an ideal $\mathbb Z_2$-register requires one qubit. For continuous Lie groups (e.g. $SU(3)$), ideal $G$-registers cannot be made with any finite number of qubits -- we must settle for approximate $G$-registers (typically with Hilbert space $\tilde{\mathcal H}_G = \operatorname{span}\{\left|g\right>\}_{g \in \tilde G}$ for some subset $\tilde G \in G$) which leads to intrinsic discretization error. The tradeoff between number of qubits and the size of discretization error will be crucial to constructing efficient simulations near-term.

In this work, we consider only the case of ideal $G$-registers. Complications arising from the discretization, which may include the breaking of gauge invariance~\cite{Stryker:2018efp} and difficulties with the definition of the kinetic and Fourier gates are not considered here. A proposal for approximate $SU(N)$ registers that does not have these problems may be found in~\cite{Hackett:2018cel,Alexandru:2019nsa}.

For the Hamiltonians we consider, a set of useful primitive gates defined on the $G$-register is: inversion, matrix multiplication, trace, and the quantum Fourier transform. The inclusion of matter fields requires additional multiplication and Fourier transform gates, and an inner product gate.  A sample implementation for these operations is specified in Appendix~\ref{app:d4-register} for the $D_4$-register case.

The inversion gate acts on a single register to produce the register of its inverse group element.  This is defined in the fiducial basis by
\begin{equation}
\mathfrak U_{-1} \left|g\right> = \left|g^{-1}\right>\text.
\end{equation}
The other group operation is matrix multiplication, which requires a two-register gate, defined by
\begin{equation}
\mathfrak U_\times \left|g\right> \left|h\right> = \left|g\right> \left|gh\right>\text.
\end{equation}
Here we have $\mathfrak U_\times$ implementing left multiplication, in the sense that the content of the target register was multiplied on the left. This is all that is required here. In general, the combination of $\mathfrak U_{-1}$ and left-acting $\mathfrak U_\times$ permits right multiplication: $\mathfrak U_{\times,R}(1,2) = \mathfrak U_{-1}(1) \mathfrak U_{-1}(2) \mathfrak U_\times(2,1) \mathfrak U_{-1}(2) \mathfrak U_\times(1,2)$.

Most quantum gatesets --- certainly all gatesets proposed for practical use --- contain the inverse of every gate as a primitive. This makes it easy, given a circuit implementing one unitary, to obtain a circuit implementing its inverse. Therefore, if $\mathfrak U_\times^\dagger$ is available, $\mathfrak U_{-1}$ is readily obtained with the assistance of an ancillary register.

The trace of a plaquette appears in our gauge Hamiltonian, and so to perform this operation we combine with the matrix multiplication gate with a single-register trace gate:
\begin{equation}
\mathfrak U_{\Tr}(\theta) \left|g\right> = e^{i \theta \Re\Tr g} \left|g\right>.
\end{equation}

In our construction, the final gate required on the $G$-register is the Fourier transform gate~\cite{coppersmith2002approximate,RevModPhys.68.733,jozsa1998quantum} $\mathfrak U_F$. This gate acts on a $G$-register to rotate into Fourier space. It is defined by
\begin{equation}
\mathfrak U_F \sum_{g \in G} f(g)\left|g \right>
=
\sum_{\rho \in \hat G} \hat f(\rho)_{ij} \left|\rho,i,j\right>
\end{equation}
The second sum is taken over $\rho$, the representations of $G$, and $\hat f$ denotes the Fourier transform of $f$. This gate diagonalizes what will be the `kinetic' part of the Trotterized time-evolution operator. After application of the gate, the register is no longer a $G$-register but a $\hat G$-register.

The $G$-register suffices for the representation of the gauge field. In order to represent scalar matter fields, we introduce the $\mathbb C^N$-register, where $N$ is the dimension of the fundamental representation\footnote{For matter in the adjoint representation, $N$ is the dimension of the adjoint representation.} of $G$. This register stores a component of $\mathbb C^N$, and we will define gates allowing transformation under elements of $G$. As with the $G$-register, it is only possible to implement a `truncated' $\mathbb C^N$ register. (For fermionic matter fields, the $\mathbb C^N$-register is unneeded, and no truncation is necessary.)

Three gates are needed to make the $\mathbb C^N$-register useful in a quantum simulation. The fundamental (or adjoint) multiplication gate $\mathfrak U_{\times,f}$ is defined by
\begin{equation}
\mathfrak U_{\times,f} \left|g\right>\left|\phi\right> = \left|g\right> \left|g\phi\right>
\text.
\end{equation}

The inner product gate, necessary for the field-theory kinetic term (sometimes called the `hopping' term, to distinguish it from the quantum mechanical kinetic term), is defined by
\begin{equation}
\left<\tilde \phi_1 \tilde \phi_2\right|\mathfrak U_{\left<\cdot,\cdot\right>}(\theta)\left|\phi_1 \phi_2\right> =
\delta^{\tilde\phi_1}_{\phi_1}
\delta^{\tilde\phi_2}_{ \phi_2}
e^{i \theta \left[\phi_2^\dagger \phi_1 + \phi_1^\dagger \phi_2\right]}
\text.
\end{equation}
The inner product is generally a complex number, but this gate extracts only the real part while remaining unitary. Happily, this is all required for simulation of field theory, where only the linear combination $\phi^\dagger_2 \phi_1+\phi^\dagger_1 \phi_2$ appear.

Lastly, the scalar Fourier transform gate is defined analogously to $\mathfrak U_F$ above, except for $\mathbb C^N$: we notate this $\mathfrak U^{\mathbb C^N}_F$. The resulting register is a $\mathbb C^N$-register. Since the quantum Fourier transform on $\mathbb C^N$ factorizes into $N$ separate transforms on $\mathbb C$, the implementation of $\mathfrak U^{\mathbb C^N}_F$ consists of $N$ separate calls to $\mathfrak U^{\mathbb C}_F$.

\section{Pure Gauge}\label{sec:yang-mills}

In this section, we construct a time-evolution operator suitable for the simulation of pure gauge theories on a quantum computer.  We take care to exactly preserve gauge invariance (up to unavoidable gate errors and noise). We take the standard approach, by discussing the Hamiltonian formulation of the theory, and then Trotterizing the Hamiltonian into a sum of pieces easily implemented via the primitive constructs of Sec.~\ref{sec:prerequisites}. This approach will be extended in Sec. \ref{sec:scalar} and \ref{sec:spinor} to include scalar and fermionic matter fields, respectively. 

A similar approach to nonabelian gauge simulation has been discussed previously~\cite{Bender:2018rdp,Zohar:2018nvl,Zohar:2012ay,Tagliacozzo:2012df,Zohar:2016iic,Zohar:2015hwa}. In this section, we take care to work only with logical qubits and gates, keeping the algorithm independent from the underlying geometry of the quantum computer. Additionally, we avoid the need for ancillary $G$-registers.

In the present work, we use the basis diagonalizing the gauge field operators $U$, rather than their conjugate momenta, unlike e.g. \cite{Byrnes:2005qx} in which the fiducial basis is chosen to diagonalize the kinetic term of the Hamiltonian. Our choice simplifies the calculation of plaquettes and Wilson lines, as well as the inclusion of matter fields. Additionally, simulations done in this `position' basis correspond more closely to calculations performed in classical lattice field theory.

We begin by reviewing the Hamiltonian formulation of lattice gauge theory~\cite{PhysRevD.11.395}. Each gauge link has a local Hilbert space $\mathbb C G$. Combined, the $L$ links on our lattice have Hilbert space $\mathcal H=\mathbb C G^{\otimes L}$.

On a lattice with $N$ sites, the space of gauge transformations is $G^N$, where one element of $G$ must be specified at each site. The transformation rule for a link $U_{ij}$ from site $i$ to site $j$ is $U_{ij} \mapsto V_j U_{ij} V^\dagger_i$. The linear action of $V \in G^N$ on the Hilbert space is given by $\phi(V)$:
\begin{equation}
\phi(V) \left|U_{ij}\hdots\right> = \left|V_j U_{ij} V^\dagger_i\hdots\right>
\end{equation}

Gauge invariance demands that two states differing only by a gauge orbit be considered physically equivalent.  The physical Hilbert space\footnote{The Gribov ambiguity prevents us from unambiguously assigning a single representative configuration to each gauge orbit; however, we do not take that approach. Physical states in the Hilbert space are not configurations, but formal linear combinations of configurations, and these can be constructed unambiguously.}, then, is $\mathcal H_P = \mathbb C G^{\otimes L} / \phi(G^{\otimes N})$.

$\mathcal H_P$ can be obtained from the larger space $\mathcal H$ by a gauge-symmetrization operator, which acts as a projection operator:
\begin{align}\label{eq:projection}
P \left|U_{12}\hdots\right> &= \frac{1}{|G|^N}\left(\int_G \d V_1 \int_G \d V_2 \hdots\right)
\left|V_2 U_{12} V^\dagger_1\hdots\right>\nonumber\\
&=
\frac{1}{|G|^N}\int_{G^N} \d V \; \phi(V) \left|U\right>
\end{align}

This discussion of the gauge-invariance of physical states is equivalent to the imposition of Gauss's law. An implementation of the operator $P$ for $U(1)$ is given by~\cite{Stryker:2018efp}; however our method does not require implementing $P$.

We define a Hamiltonian which acts on the entire space $\mathcal H$ of the gauge theory, implicitly defining a `physical' Hamiltonian on the subspace $\mathcal H_P$. As will be made clear in later sections, this choice of Hamiltonian is motivated by a desire to recover the Euclidean Wilson action.
\begin{align}\label{eq:hamiltonian}
H =&
-\frac{1}{g^2 a} \sum_p \Re \Tr \left[\prod_{\left<ij\right>\in p} U_{ij}\right]
+
\frac{4 g^2}{a} \sum_{\left<ij\right>} \pi_{ij}^2\nonumber\\
=& H_V+H_K
\end{align}
The first term, $H_V$, is a sum over spatial plaquettes $p$ and can be thought of as a potential term. The trace is taken in a suitable representation of the gauge group $G$ --- for matrix groups, this is typically the fundamental representation. The second term, $H_K$, is the quantum-mechanical kinetic term describing the mechanics of a free particle moving on the surface of $G$. Therefore, $\pi_{ij}^2$ is the Laplace-Beltrami operator on the surface $G$, and $\pi_{in}$ is the momentum operator conjugate to $U_{ij}$. This prescription gives us no guidance for how to form the kinetic term of a discrete gauge theory --- this is addressed via the transfer matrix formalism in Section~\ref{sec:path-integral}.

Not only is the Hamiltonian gauge-invariant, but \emph{each term} is gauge-invariant. This is crucial, as it allows us to Trotterize without breaking gauge invariance at any order in $\Delta t$.  A time evolution operator suitable for implementation on a quantum computer is then
\begin{equation}
\mathcal U(t) = \prod_{t / \Delta t} e^{-i H_K \Delta t} e^{-i H_V \Delta t}\text. 
\end{equation}
Below, we will use the notation $\mathcal U(\Delta t)$ to indicate a single $\Delta t$ step.  We now describe how the operators $e^{-i H_K \Delta t}$ and $e^{-i H_V \Delta t}$ may be obtained. First, each part (kinetic and potential) of the Hamiltonian consists of mutually commuting, local terms, and each such term may be treated individually and sequentially without approximation. That is,
\begin{equation}
e^{-i H_K \Delta t}
=
\prod_{\left<ij\right>} \mathcal U_K^{(1)}(i,j)
\text{ and }
e^{-i H_V \Delta t}
=
\prod_p \mathcal U_V^{(1)}(p)\text,
\end{equation}
Here, again because the factors commute, the order of application of the $\mathcal U_V^{(1)}$ makes no difference to the final result.
The plaquette evolution $\mathcal U_V^{(1)}$ is constructed by first preparing the product of the plaquette in an ancillary register, and then operating on the ancillary with $\mathfrak U_{\Tr}(\frac 1 {g^2 a})$. $\mathcal U_K^{(1)}$ is implemented by diagonalizing via $\mathfrak U_F$, and then applying a diagonal unitary:
\begin{equation}
\mathcal U_K^{(1)}(i,j) =
\mathfrak U_F
\mathfrak U_{\mathrm{phase}}
\mathfrak U_F^\dagger.
\end{equation}

\begin{figure*}
\includegraphics{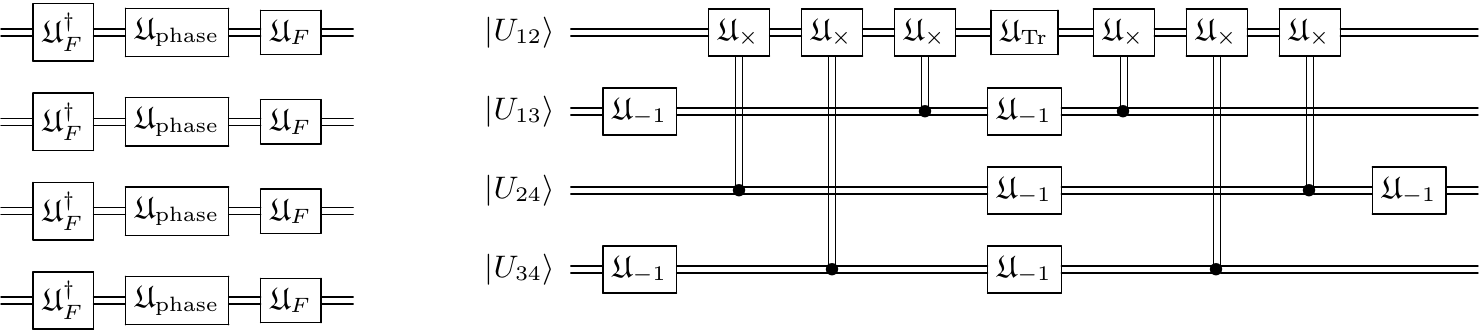}
\caption{Circuits for the propagation of a pure-gauge lattice field theory. The first circuit implements $\mathcal U_K^{(1)}$ on four links (in general, $L$ links are needed). The second circuit shows the application of $\mathcal U_V^{(1)}$ to a single plaquette $\Re \Tr U_{13}^\dagger U_{34}^\dagger U_{24} U_{12}$, and must be applied to every plaquette in the theory. Note that in these circuits, we use a doubled line to represent a $G$-register, rather than a single qubit.\label{fig:gauge-circuits}}
\end{figure*}

Circuits implementing $\mathcal U_K^{(1)}$ and $\mathcal U_V^{(1)}$ are shown in \fig{gauge-circuits}. The implementation of $\mathfrak U_{\mathrm{phase}}$ depends strongly on the group and the implementation of the $G$-register, but in all cases it is a diagonal operator. The total gate requirements for the propagation are shown in Table~\ref{tab:cost}.

Note that $\mathcal U_V$ involves multiple $\mathcal U_V^{(1)}$, some of which affect overlapping links. Because all potential operators commute (indeed all $\mathcal U_V^{(1)}$ are diagonal in the fiducial basis), the order they are performed in does not matter.

\begin{table}
\begin{center}
\begin{tabular}
{c| c}
\hline\hline
Gate & Number\\
\hline
$\mathfrak U_F$ & $2L \frac {T}{\Delta t}$ \\
$\mathfrak U_{\text{phase}}$ &$L \frac {T}{\Delta t}$ \\
$\mathfrak U_{-1}$ & $6 N_P \frac {T}{\Delta t}$ \\
$\mathfrak U_\times$&$6 N_P \frac {T}{\Delta t}$\\
$\mathfrak U_{\Tr}$&$N_P \frac {T}{\Delta t}$\\
\hline\hline
\end{tabular}
\end{center}
\caption{\label{tab:cost}Gate requirements for the propagation of a lattice with $N_P$ plaquettes and $L$ links, for a time $T$ with time-steps of size $\Delta t$}
\end{table}
When preparing an initial state, it is important to ensure that it lies within the physical subspace $\mathcal H_P$. One procedure for projecting onto the gauge-invariant states for an Abelian gauge theory is provided in \cite{Stryker:2018efp}.

Although the preparation of the ground state $\left|\Omega\right>$ is beyond the scope of this paper, it is worth discussing that common proposals typically require the time-evolution operator $\mathcal U(t)$ as constructed in this section. Assuming such methods are used, nonequilibrium physics can be computed by preparing the ground state with $\mathcal U(t)$, and then evolving with a different time-evolution operator $\mathcal {\tilde U}(t)$~\cite{Lamm:2018siq}. A typical expectation value of interest, accessible in such a scheme, is 
\begin{equation}
E(t) = \left<\Omega\right|\mathcal {\tilde U}^\dagger(t) \hat E \mathcal {\tilde U}(t)\left|\Omega\right>
\end{equation}
where $\hat E$ measures the expectation of a plaquette (i.e. the energy density of the gauge field). Other expectation values of interest are discussed in Sec.~\ref{sec:correlators} below.

\section{The Lattice Path Integral}\label{sec:path-integral}
Substantial insight into QCD has been achieved via Euclidean-spacetime lattice methods on classical computers. Accordingly, the Euclidean lattice path integral occupies a dominant place in the QCD literature and the minds of practitioners. This is in contrast to quantum simulations where the Hamiltonian formulation appears more natural. In this section, we connect the Hamiltonian formalism to the Euclidean lattice path integral via a transfer matrix. This implies an important practical consequence: the determination of appropriate bare coupling constants may be performed on a classical computer. This procedure also guides us in constructing a Hamiltonian for a discrete gauge theory (discussed in \cite{Harlow:2018tng}).

Our formalism for the transfer matrix, $T$, slightly differs from the usual one~\cite{Luscher:1976ms,Mitrjushkin:2002nk,Creutz:1999zy,Caracciolo:2012qw,Creutz:1976ch} in that $T$ is defined on the entire space $\mathcal H$. The usual transfer matrix is defined only on the physical Hilbert space, and may be obtained by projection with $P$. In the course of performing this projection, the lattice path integral with the Wilson action is recovered.

The transfer matrix is an approximation to imaginary-time evolution by one unit of time: $T \approx e^{- H}$.
We construct the transfer matrix in the fiducial basis of $\mathcal H$ from separate kinetic $T_K$ and potential $T_V$ contributions via the split-operator approximation
\begin{equation}\label{eq:transfer}
T = T_V^{1/2} T_K T_V^{1/2}
\end{equation}
where the potential term resembles the product of spatial plaquettes as appears in the Wilson action
\begin{equation}
\left<\tilde U_{12}\hdots\right|T_V\left|U_{12}\hdots\right> =
\delta_{U_{12}\cdots}^{\tilde U_{12}\cdots}
\exp\left(\frac {a_0} {a g^2}\sum_x W_{\mu\nu}(x)\right)\text.
\end{equation}
Here $a$ is the spatial lattice spacing, and $a_0$ the temporal spacing; we do not assume an isotropic lattice. We have borrowed from lattice field theory the Wilson plaquette
\begin{equation}W_{\mu\nu}(x) = \Re \Tr [U^\dagger_{x,x+\hat\nu} U^\dagger_{x+\hat\nu,x+\hat\mu+\hat\nu} U_{x+\hat\mu,x+\hat\mu+\hat\nu} U_{x,x+\hat\mu}],
\end{equation}
and $\mu,\nu$ are restricted to space-like directions.

For the kinetic term there is no coupling between different links, but it is related in lattice field theory to plaquettes containing timelike links:
\begin{align}
\langle\tilde U_{12}\hdots|T_K|U_{12}\hdots\rangle =
\prod_{\left<ij\right>} e^{\frac a {g^2 a_0}\Re \Tr \left[\tilde U_{ij}^\dagger U_{ij}\right]}
\end{align}
 Note that, like the Trotterized real-time evolution, $[T,P]=0$ due to the fact that $T_V$ and $T_K$ individually commute with $P$.

When the transfer matrix is restricted to physical states, it exactly reproduces the usual lattice path integral, with the Wilson action.
\begin{align}\label{eq:transfer-partition}
Z &= \Tr_P T^\beta = \Tr (PTP)^\beta\nonumber\\
&=
\left(\int_G \d U_{12}\cdots\right) \exp\left[\frac 1 {g^2} \sum \Re \Tr \prod U_{ij}\right]
\end{align}
Here $\Tr_P$ denotes the trace over only the physical subspace $\mathcal H_P$. The inverse temperature is denoted $\beta$, and $Z$ is the Euclidean lattice approximation to the partition function.

Now that we have a transfer matrix corresponding to the Wilson action, we may connect it to the Hamiltonian. This is done in detail by Creutz \cite{Creutz:1976ch} for the gauge-fixed transfer matrix, so we will only summarize the gauge-free result (which is essentially the same) here. The potential part $T_V$ of the transfer matrix is exactly $e^{-H_V}$. The relationship between the kinetic part $H_K$ of the Hamiltonian and $T_K$ is more subtle. It is shown in \cite{Creutz:1976ch} that, in the limit where the temporal lattice spacing $a_0$ is taken to $0$, the kinetic part of the Hamiltonian is recovered.


With a mapping between the Hamiltonian and lattice action through the transfer matrix, the bare constants in \eq{hamiltonian} for a desired renormalized, physical set of couplings are the same as the lattice action ones. This is particularly important for ensuring that Lorentz invariance is recovered in the continuum limit, but also allows us to perform simulations at known values of physical parameters. In particular, in a lattice QCD simulation, the task of determining appropriate bare parameters for a desired physical pion mass may be performed cheaply on a classical computer, reserving the quantum processor for the difficult real-time evolution.

Clearly, the Hamiltonian evolution on a quantum computer is closely related to standard classical lattice simulations. Indeed, the two differ only in: the discretization of the gauge group (quantum computers have smaller $G$-registers), whether the Trotterization is done in real or imaginary time, and the exact form of the kinetic term. The first difference can be removed by performing a classical simulations with reduced precision, and the others are reduced at small temporal lattice spacing.

In the case of the Trotterization $\Delta t$ being equal to $a$, the quantum evolution is simply an analytic continuation of the usual isotropic lattice calculations performed in lattice QCD. This implies that Lorentz invariance is recovered in the continuum limit.

We conclude by exploiting this procedure to produce a Hamiltonian for a discrete gauge theory --- that is, a gauge theory where the gauge group is a discrete group (such as $D_4$, demonstrated below), rather than a connected manifold. In this case, the kinetic part of the Hamiltonian is determined not by a Laplacian on a manifold, but can be derived from relevant portion of the transfer matrix by taking a logarithm.
\begin{align}
H = H_V& + H_K\nonumber\\
\text{where }
H_V = -\log T_V &\text{ and } H_K = -\log T_K\text,
\end{align}
where $T_V$ and $T_K$ are defined exactly as in \eq{transfer}.
Note that $H_K$ is only well-defined if $T_K$ is positive-definite: this fact is established for a general discrete group (with the Wilson action) in Appendix \ref{app:positive-definite}.

As a consequence of Schur's lemma, the kinetic transfer gate defined here is diagonal in the Fourier basis. This implies that, where $H_K$ is obtained in this fashion, $\mathcal U_K$ may be easily implemented from $\mathfrak U_F$.

\section{Correlators}\label{sec:correlators}
Having constructed time-evolution for a pure gauge theory, we now must show how observables of physical interest may be obtained. In this section, we show how to obtain a plaquette-plaquette correlator and the expectation value of an arbitrary, temporally extended Wilson loop. Such correlation functions serve as the nonperturbative input to larger calculations, such as the determination of parton distribution functions.

 We accomplish both correlators by the technique set out in \cite{PhysRevLett.113.020505}. To begin, we recall how one can compute an expectation value of a unitary operator $U$ \cite{Ortiz:2000gc} in a given state $\left|\Psi\right>$. Introducing a single ancillary qubit, we construct a controlled unitary operator $U_C$, defined by
\begin{equation}
U_C\left|\Psi\right>\left|0\right> = \left|\Psi\right>\left|0\right>
\text{ and }
U_C\left|\Psi\right>\left|1\right> = U\left|\Psi\right>\left|1\right>
\text.
\end{equation}
(Standard quantum gatesets make the construction of this operator straightforward.)

Generally, the expectation value of $U$ has both real and imaginary parts. These  must be measured separately. A measurement of $\Re \left<\Psi\right| U \left|\Psi\right>$ is given by initializing the ancillary qubit to $|+\rangle\equiv\frac{1}{\sqrt 2}\left(\left|0\right> + \left|1\right>\right)$, performing evolution via $U_C$, and then measuring $\sigma_x$ on the ancillary qubit.
\begin{equation}
\left(\langle\Psi|\otimes \langle +| \right)U_C^\dagger
\left(\id\otimes \sigma_x\right)
U_C \left(\left|\Psi\right> \otimes |+\rangle \right)
=
\Re \left<\Psi\right|U\left|\Psi\right>
\end{equation}
For $\Im \left<\Psi\right|U\left|\Psi\right>$ one performs the same process except ending with a measurement of $\sigma_y$.


With this procedure in mind, we can show how to compute a correlator of the form
\begin{equation}\label{eq:glueball-propagator}
\left<\Psi\right|\mathcal U(-t)W_{\mu'\nu'}(x')\mathcal U(t)W_{\mu\nu}(x)\left|\Psi\right>\text.
\end{equation}
In QCD this Wilson plaquette correlator corresponds to a glueball propagator. The operator in \eq{glueball-propagator} is clearly not unitary, so it cannot be directly evaluated by means of the procedure described above. Instead, using the Hermiticity of $W_{\mu\nu}(x)$, we construct a parameterized family of unitary operators. The derivatives of this family give the operator of interest. Explicitly, this is done via a time-dependent perturbation of the Hamiltonian:
\begin{equation}
H_{\epsilon_1,\epsilon_2}(\tau) = H_0 +
\epsilon_2 \delta(\tau - t) W_{\mu'\nu'}(x')
+
\epsilon_1 \delta(\tau) W_{\mu\nu}(x)
\end{equation}
Time evolving forward in time with $H_{\epsilon_1,\epsilon_2}$, and back with $H_0$, allows us to measure the expectation value $C(\epsilon_1,\epsilon_2) \equiv \left<\Psi\right|\mathcal U(-t) \mathcal U_{\epsilon_1,\epsilon_2}(t)\left|\Psi\right>$. Differentiating twice with respect to the perturbation yields the plaquette-plaquette correlator as desired.
\begin{equation}
-\left.\frac{\partial^2 C(\epsilon_1,\epsilon_2)}{\partial \epsilon_1\partial \epsilon_2}\right|_{\epsilon_1 = \epsilon_2 = 0}
= 
\left<\mathcal U(-t)W_{\mu'\nu'}(x')\mathcal U(t)W_{\mu\nu}(x)\right>
\end{equation}
In practice, this derivative must be obtained by finite differencing, after computing $C(\epsilon_1, \epsilon_2)$ for several small values of $\epsilon$. This approach is naturally extended to the computation of $k$-time correlation functions, at the cost of requiring the $k$th numerical derivative.

The terms $W_{\mu\nu}(x)$ added in the perturbation also appear in the original Hamiltonian, and are readily implemented with the primitive gates defined in Section~\ref{sec:prerequisites}.

We can use the same method to compute the expectation values of temporally-extended Wilson loops, $\left<\Re \Tr U^\dagger_{ij}(t) U_{ij}(0)\right>$.  To do so, we decompose it into terms $\left<\Re ([U^{\dagger}_{ij}(t)]^{ba} [U^{ab}_{ij}(0)]^{ab})\right>$ where $a,b$ are color indices. We can use this to derive a perturbed Hamiltonian
\begin{align}
H^{ab}_{\epsilon}(\tau) = H_0
&+ \epsilon_2 \delta(\tau - t) \Re [U^\dagger_{ij}]^{ba}
+ \epsilon_1 \delta(\tau) \Re [U_{ij}]^{ab}\nonumber\\
&+ \tilde \epsilon_2 \delta(\tau - t) \Im [U^\dagger_{ij}]^{ba}
+ \tilde \epsilon_1 \delta(\tau) \Im [U_{ij}]^{ab}
\text,
\end{align}
which can be utilized in a correlator $C^{ab}(\epsilon_1,\tilde\epsilon_1,\epsilon_2,\tilde\epsilon_2) \equiv \left<\Psi\right|\mathcal U(-t) \mathcal U^{ab}_{\epsilon}(t)\left|\Psi\right>$ 
We now take the difference of two second derivatives of the resulting correlator:
\begin{align}
- \left[
\frac{\partial^2}{\partial \epsilon_1 \partial \epsilon_2}
-
\frac{\partial^2}{\partial \tilde \epsilon_1 \partial \tilde \epsilon_2}
\right]_{\epsilon=0}
&C^{ab}(\epsilon_1,\epsilon_2,\tilde\epsilon_1,\tilde\epsilon_2)
=\nonumber\\
&\Re \left<
\mathcal U(-t) [U_{ij}^\dagger]^{ba} \mathcal U(t) [U_{ij}]^{ab}
\right>
\text.\label{eq:wilson-loop-correlator}
\end{align}
Summing over the $a,b$ yields the desired, gauge-invariant trace. The imaginary part can be obtained similarly.

For many gauge groups, the different correlators $\left<(U^\dagger_{ij}(t))^{ba} (U_{ij}(0))^{ab}\right>$ are related by gauge symmetry. Consequently, for these gauge groups, the correlator in \eq{wilson-loop-correlator} need only be evaluated for one particular selection of $a$ and $b$, as all the other terms will be equal. In particular, this is true for $D_4$ (simulated below) and $SU(N)$ in the fundamental representation.

The perturbation introduced, unlike the one used for plaquette-plaquette correlators above, is \emph{not} gauge-invariant, meaning that the state during this time evolution does not lie within the physical subspace $\mathcal H_P$.
Also unlike the plaquette-plaquette perturbation, this perturbation contains terms not present in the original Hamiltonian, and requires gates not specified in Section~\ref{sec:prerequisites}. The gates required are diagonal (in the fiducial basis) phase gates on individual $G$-registers.

This method can also compute a non-gauge-invariant expectation value, e.g. $\left<\Tr U_{ik}(t) U_{ij}(0)\right>$ where $j \ne k$. Provided the initial state is correctly prepared, this expectation value must be $0$.

\section{Scalar fields}\label{sec:scalar}
Here we extend our Hamiltonian construction to include scalar matter fields which transform in a $D$-dimensional representation of $G$. The Hilbert space defined in Section \ref{sec:yang-mills} is extended via tensor product to include $D$ scalar degrees of freedom at each site. The new Hilbert space is $\mathcal H = (\mathcal H_G)^{\otimes L} \otimes (\mathcal H_S)^{\otimes N}$, where $\mathcal H_G$ is the one-link Hilbert space defined above, $N$ is the number of sites, and $\mathcal H_S$ is the one-site Hilbert space. For $\mathcal H_S$, it is convenient to work in the basis of occupation number, writing $\mathcal H_S = \mathop{\mathrm {span}}\{\left|n_1\cdots n_D\right> : n_i = 0,1,\ldots\}$.  This is different from classical lattice field theory where scalars are typically represented in a basis diagonalizing the field operator or integrated out, but is more natural for quantum simulations.

As before, we first consider how gauge transformations act on $\mathcal H$. A gauge transformation is still specified by $V_i \in G$ at each site $i$. The gauge links still transform as $U_{ij} \mapsto V_j U_{ij} V_i^\dagger$, and the scalar fields transform (in the fundamental) via $\phi_i \mapsto V_i \phi_i$. In this way, $\phi_j^\dagger U_{ij} \phi_i$ is gauge-invariant. The action of a gauge transformation $V$ on the Hilbert space is
\begin{equation}
\left|U_{12}\cdots\right> \otimes \left|\phi_1\cdots\right>
\mapsto
\left|(V_2 U_{12} V_1^\dagger)\cdots\right>\otimes \left|(V_1\phi_1)\cdots\right>
\text.
\end{equation}
The gauge-symmetrization operator is still defined by \eq{projection}, but acts on the expanded Hilbert space.

To represent the physical Hilbert space on a quantum computer, we now make use of the $\mathbb C^N$-registers transforming under $G$. For the scalar fields we consider, the Hamiltonian is
\begin{align}
H =&
\beta \sum_p \Re \Tr \left[\prod_{\left<ij\right>\in p} U_{ij}\right]
+
\frac 1 2 \sum_{\left<ij\right>} \pi_{ij}^2
\nonumber\\&+
\frac 1 2 \sum_i \nabla^2{\phi_i} + 
\frac{m^2}{2} \sum_i \phi_i^\dagger \phi_i
+
\frac 1 2
\sum_{\left<ij\right>} \left|\phi_j - U_{ij} \phi_i\right|^2.
\end{align}
Although we assume scalar interactions only to order $\phi^2$, the generalization to higher-order couplings is trivial and introduces no complications into what follows.  We construct a Trotterized time evolution operator, just as in Section \ref{sec:yang-mills}, by decomposing into potential and kinetic terms.  The new gates required for the scalar fields are $\mathcal U_{M}^{(1)}$ (implementing the mass term), $\mathcal U_H^{(1)}$ (implementing the hopping term), and $\mathcal U^{(1)}_{K,\mathrm{scalar}}$.  These are defined by
\begin{align}
\mathcal U_{M}^{(1)}(i) &= e^{-i \Delta t \left(d+\frac {m^2}{2}\right) \phi_i^\dagger \phi_i}
\text,
\\
\mathcal U_{H}^{(1)}(i,j) &=
e^{-i \Delta t \Re \left(\phi_j, U_{ij} \phi_i\right)}
\text{, and}
\\
\mathcal U^{(1)}_{K,\mathrm{scalar}}(i,j)
&= e^{-i \Delta t \nabla^2}
\text.
\end{align}
Note that the mass and hopping gates commute, as both are diagonal in the fiducial basis. The same-site $\phi_i^\dagger \phi_i$ terms generated by the hopping term of the Hamiltonian have been absorbed into the mass gate, assuming a square lattice of dimension $d$. These gates, together with the pure-gauge $\mathcal U_{V,K}^{(1)}$, are combined to form
\begin{align}
\mathcal U_V &=
\prod_p \mathcal U_{V,\mathrm{gauge}}^{(1)}(p)
\prod_i \mathcal U_{M}^{(1)}(i)
\prod_{i,j} \mathcal U_{H}^{(1)}(i,j)\\
\mathcal U_K &=
\bigotimes_{\left<ij\right>} \mathcal U_{K,\mathrm{gauge}}^{(1)}(i,j)
\otimes
\bigotimes_{i} \mathcal U_{K,\mathrm{scalar}}^{(1)}(i).
\end{align}
With these, the full time-evolution operator for a single time-step is
\begin{align}
\mathcal U(\Delta t)&= \mathcal U_V \mathcal U_K
\end{align}

A single time step, therefore, consists of five steps on the quantum computer:
\begin{enumerate}
\item For each plaquette, compute the plaquette product and apply $\mathfrak U_{\Tr}(\Delta t / g^2 a)$. This implements $\mathcal U_{V,\mathrm{gauge}}$.
\item For each link, apply the scalar hopping gate $\mathcal U_{H}^{(1)}$.
\item For each site, apply the scalar mass gate $\mathcal U_M^{(1)}$.
\item For each link, apply the kinetic gate $\mathcal U_{K,\mathrm{gauge}}^{(1)}$.
\item For each site, apply the scalar kinetic gate $\mathcal U_{K,\mathrm{scalar}}^{(1)}$.
\end{enumerate}
This is the general algorithm for time-evolving gauge fields in the presence of matter.  Before moving on to discuss scalar field correlators, we note that the transfer matrix discussion of Sec.~\ref{sec:path-integral} could be repeated here, including the scalar fields, to connect the evolution of the quantum computer to a classical, Euclidean spacetime path integral. Thus, renormalization calculations again may be performed classically.

Steps 1 and 4 above are the same as for a pure-gauge simulation. Step 2 requires two calls to $\mathfrak U_{\times,f}$, two calls to $\mathfrak U_{-1}$, and one call to $\mathfrak U_{(\cdot,\cdot)}$. Step 3 is a register-implementation dependent diagonal phase gate. Step 5, analogous to the implementation of $\mathcal U_{K,\mathrm{gauge}}^{(1)}$,  entails $2N$ calls to $\mathfrak U_F^{\mathbb C}$ and $N$ diagonal phase gates per site, where $N$ is the dimension of the scalar's representation.

The procedure for computing plaquette correlators is the same as in Section \ref{sec:correlators}.

The construction of Section~\ref{sec:correlators} may be generalized to obtain the gauge-invariant two-point function of a fermion, connected by a Wilson line: $\left<\bar\psi(y,t) W \psi(x,0)\right>$. However, for a general Wilson line, this construction requires one derivative to be taken at every timeslice. As the derivatives must be taken via finite differencing, this is not a plausible method for computing these correlators, even in the far future. Constructing a more efficient method remains an open problem.

\section{Fermion fields}\label{sec:spinor}
Here we consider the inclusion of spinor matter fields, which can be done in the same manner as was discussed above for scalar fields. Also as before, the representation of fermionic fields on the quantum computer will not match that on a classical computer, where they are typically integrated out. In the implementation of spinor fields on a quantum computer, while the register is trivial to specify and exact, a significant technical difference emerges: the fundamental operators provided on most digital quantum computers are bosonic, meaning that operators acting on different qubits will commute with each other. For spinor fields, we want fermionic operators, which anticommute at different physical sites. We will translate commuting operators into anticommuting operators via the Jordan-Wigner transformation, paying the price that local fermionic operators are generated by nonlocal bosonic operators.  More efficient, albeit complex, schemes, relating local fermionic operators to more local bosonic operators have been devised~\cite{Verstraete:2005pn,Zohar:2018cwb,2016PhRvA..94c0301W}.

The Hilbert space of a gauge theory with $d$-component fermions can be written
$\mathcal H = \left(\mathcal H_G\right)^{\otimes L} \otimes \left(\mathcal H_F\right)^{\otimes dN}$, where there are $L$ links, $N$ sites, and $\mathcal H_F = \operatorname{span}\{\left|0\right>,\left|1\right>\}$ is the Hilbert space of a single fermion on a site\footnote{As a consequence of writing the fermionic space as a tensor product, the fermionic operators will be nonlocal.}. Note that, as a consequence of the exclusion principle, $\mathcal H_F$ can be put on the quantum computer without approximation or truncation.

Writing the Hamiltonian using operators that obey the standard fermionic anticommutation relations is easily accomplished in the absence of gauge fields via the Jordan-Wigner transformations~\cite{Jordan:1928wi}: each fermionic operator pair is given an index from $1\cdots dN$, and the $i$th operators are constructed from bosonic operators $a_i$ and $a^\dagger_i$:
\begin{equation}
\psi_i = \sigma^{(z)}_1 \cdots \sigma^{(z)}_{i-1} a_i
\text{ and }
\psi^\dagger_i = \sigma^{(z)}_1 \cdots \sigma^{(z)}_{i-1} a^\dagger_i
\text.
\end{equation}
In the presence of gauge fields, we must ensure that the fermionic operators obey the appropriate gauge transformation law. To be concrete, consider when the fermions transform in the fundamental of $SU(3)$. Then at each site $i$, for each spinor index $\alpha$, there are three annihilation and creation operators, which must transform under gauge transformations as
\begin{equation}
\psi^a_{\alpha i}
\mapsto
\sum_{b=1}^3 g_{ab} \psi^{b}_{\alpha i}
\end{equation}
where $a,b$ are color indices. There are now three times as many fermion operators, as for each site and each spinor, there must be three different fermionic degrees of freedom.

This may be accomplished if the bosonic operators $a,a^\dagger$ from which the fermions are constructed transform in the fundamental of $SU(3)$. Thus each operator has a color index, and $a^a \mapsto g_{ab} a^b$. Crucially, no change to the Jordan-Wigner procedure is needed since the $\sigma$-matrices are independent of the gauge group (however the $\sigma$ matrices do inherit the color index).

With fermionic operators constructed, the time-evolution operator is constructed in a similar fashion to that from Section \ref{sec:scalar} above.
\begin{equation}
\begin{split}
\mathcal U(1)&= \mathcal U_V \mathcal U_K
\\
\text{where }
\mathcal U_V &=
\prod_p \mathcal U_{V,\mathrm{gauge}}^{(1)}(p)
\prod_{i,j} U_{V,\mathrm{spinor}}^{(1)}(i,j)
\\
\text{and }
\mathcal U_K &=
\bigotimes_{\left<ij\right>} \mathcal U_{K,\mathrm{gauge}}^{(1)}(i,j)
\otimes
\bigotimes_{i} \mathcal U_{K,\mathrm{spinor}}^{(1)}(i)
\text,
\end{split}
\end{equation}
The key difference is that the spinor kinetic and potential operators are derived from the Dirac equation. The governing Hamiltonian is now
\begin{align}
H =&
\beta \sum_p \Re \Tr \left[\prod_{\left<ij\right>\in p} U_{ij}\right]
+
\frac 1 2 \sum_{\left<ij\right>} \pi_{ij}^2
\nonumber\\&+
m \sum_i \bar\psi_i\psi_i
+
\sum_{\left<ij\right>} \bar\psi_j U_{ij} \psi_i
\end{align}
and the single-site and single-link evolution operators are defined as before:
\begin{align}
\mathcal U_{V,\mathrm{spinor}}^{(1)}(i) &= e^{-i \Delta t m \bar\psi_i \psi_i}\nonumber\\
\mathcal U^{(1)}_{K,\mathrm{spinor}}(i,j) &= e^{-i \Delta t \bar\psi_j U_{ij} \psi_i}
\end{align}
An implementation of $\mathbb Z_2$ fermions is demonstrated in Section \ref{sec:demonstration}.

\section{Demonstration}\label{sec:demonstration}

In this section we present two concrete demonstrations of the method described above: first, a two-plaquette theory with the discrete, non-Abelian gauge group $D_4$, and second, a $\mathbb Z_2$ gauge field coupled to fermions. For all our simulations, we work in units where the lattice spacing is $a=1$. We perform classical simulations of a quantum computer using \texttt{qiskit}~\cite{santos2017ibm,cross2017open} in order to demonstrate the correctness of the implementation. These calculations are performed without modeling of realistic noise sources, corresponding to a perfect, error-free quantum computer.  As will be discussed below, the number of entangling gates required for a single Trotterization step is comparable to the total number used in state-of-the-art quantum simulations~\cite{nam2019ground}, and attempting to run on a current quantum computer or using a realistic noise model would be impractical.    

\begin{figure}
\centering
\includegraphics{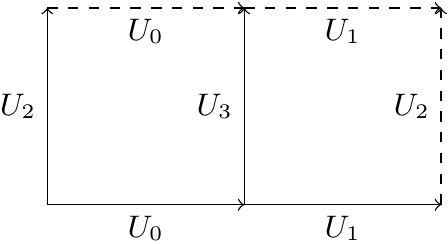}
\caption{\label{fig:lattice}The lattice geometry used for the $D_4$ gauge simulation. The plaquettes are given by $U_2^\dagger U_0^\dagger U_3 U_0$ and $U_3^\dagger U_1^\dagger U_2 U_1$.  Dash lines are used to indicate repeated links due to the periodic boundary conditions.}
\end{figure}

We simulate the $D_4$ gauge field on the two-dimensional lattice shown in \fig{lattice}. This lattice represents the smallest two-dimensional lattice which cannot be reduced to a one-dimensional theory. The simulation requires a four $D_4$ registers, and uses a total of 14 qubits: 12 for physical degrees of freedom, and 2 ancillary qubits. Note that, for brevity, we have broken with the notation of previous sections, in referring to a link not by the source and sink sites, but instead with a single direct index $0\ldots 3$.

We define a trace on $D_4$ (not uniquely specified by the group structure) by embedding $D_4$ into $U(2)$, and defining the trace via the fundamental representation of that Lie group. The embedding of $D_4 < U(2)$ is generated by the elements $\sigma_x$ and $i \sigma_z$. The lattice action for this model is
\begin{align}\label{eq:d4-action}
S =&
- \frac 1 {g^2} \sum_t \bigg(
\Re \Tr \left[U_2^\dagger(t) U_0^\dagger(t) U_3(t) U_0(t)\right]
\nonumber\\&+ \Re \Tr \left[U_3^\dagger(t) U_1^\dagger(t) U_2(t) U_1(t)\right]
\bigg)\nonumber\\
&- \frac 1 {g^2}
\sum_{P_T} \Re \Tr P_T
\text,
\end{align}
The last term is a sum over all temporal plaquettes on the $2+1$ lattice. The resulting Hamiltonian is
\begin{align}
H =&
\Re \Tr \left[U_2^\dagger(t) U_0^\dagger(t) U_3(t) U_0(t)\right]
\nonumber\\&+ \Re \Tr \left[U_3^\dagger(t) U_1^\dagger(t) U_2(t) U_1(t)\right]
\nonumber\\
&-
\sum_{i=0..3} \log T_K^{(1)}(i)
\end{align}
where $\log T_K^{(1)}$ is the one-link kinetic term, determined as discussed in Section~\ref{sec:path-integral}.

\begin{figure}
\includegraphics[width=\linewidth]{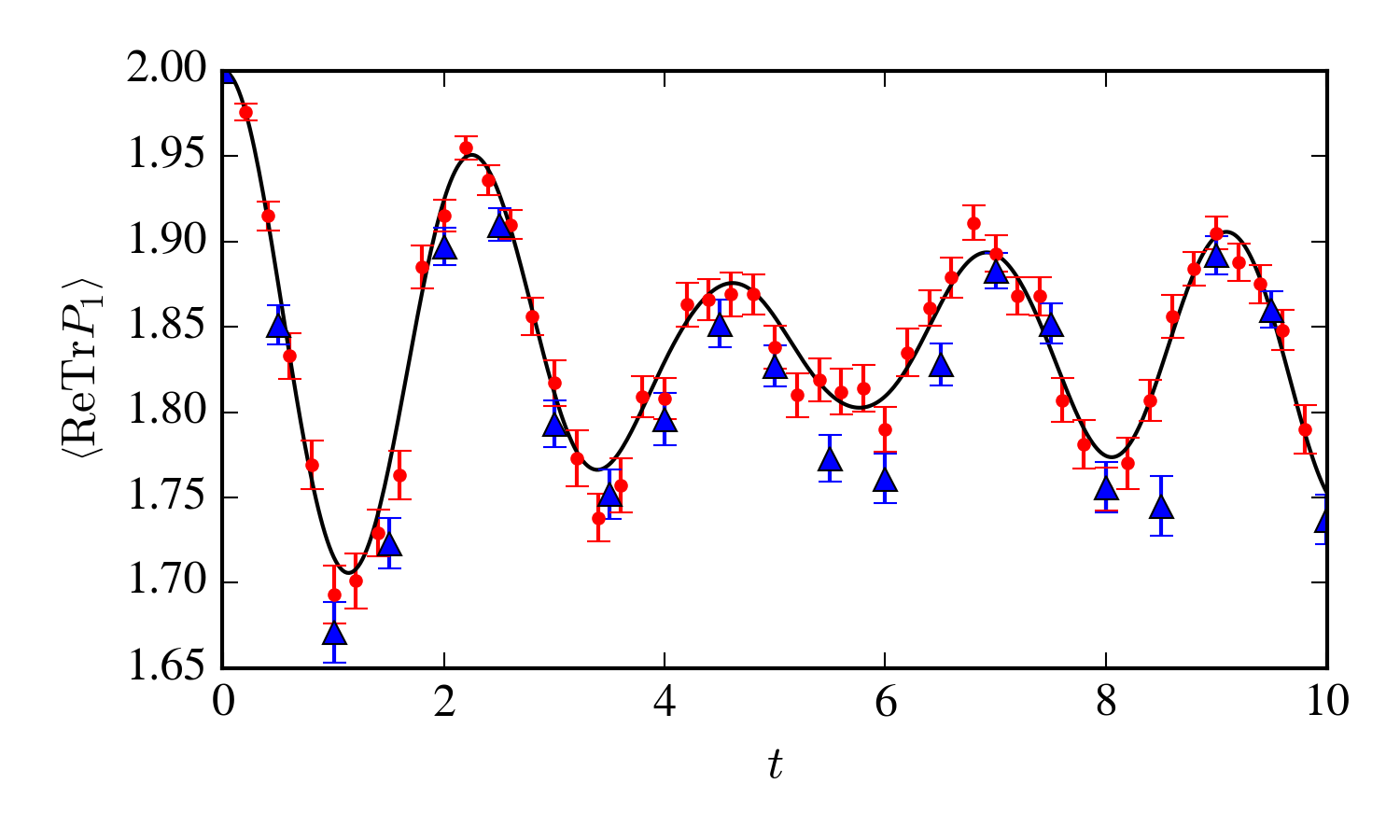}
\caption{Simulation of a $D_4$ gauge theory with two plaquettes. The expectation value of one of the plaquettes is shown as a function of $t$. The exact result is shown in black; sampled data with $\Delta t = 0.2$ ($\Delta t = 0.5$) are shown in red (blue). Differences between the simulated quantum calculation are due to sampling error (estimated with error bars) and Trotterization.\label{fig:d4-results}}
\end{figure}

The starting state of our $D_4$ simulation is the gauge-invariant projection of the eigenstate of the operators $U_i$, where all links have been assigned to the identity matrix. We evolve the system for $t=10$ with two different Trotterization time steps, one of $\Delta t = 0.2$ and one of $\Delta t = 0.5$.  With the larger Trotterization, errors build up more quickly, becoming noticeable at large times. The circuits used are detailed in Appendix \ref{app:d4-register}. In total, the quantum simulation of \fig{d4-results} entailed $\sim 200$ entangling gates per Trotterization time step. This is roughly the resources recently used to compute the ground state energy of the water molecule~\cite{nam2019ground}.   \fig{d4-results} shows the average plaquette energy of the left plaquette, as a function of time.  The only systematic error is Trotterization, as there is no noise on the simulated quantum computer.

\begin{figure}
\includegraphics[width=\linewidth]{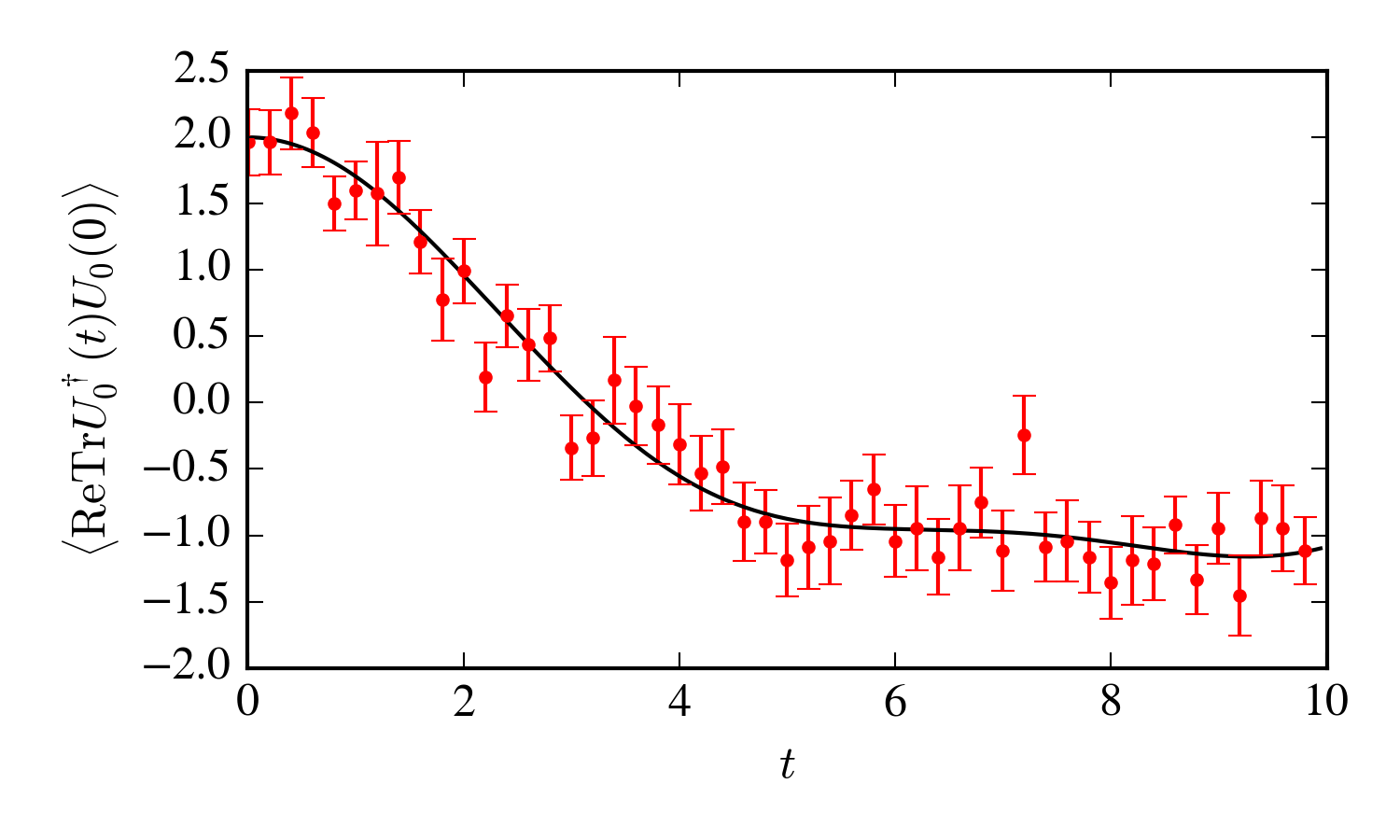}
\caption{Expectation value of a temporal Wilson loop in $D_4$ gauge theory, as a function of the time extent of the loop. For each data point, a total of $2 \times 10^5$ measurements were collected.
The systematic errors (not shown) present are Trotterization and finite differencing.
\label{fig:plot-wilson}
}
\end{figure}

For the same lattice gauge theory, we also demonstrate the procedure described in Section~\ref{sec:correlators} for determining the expectation value of a Wilson loop with temporal extent $t$: $\left<\Re \Tr U_{0}^\dagger(t) U_0(0)\right>$. Simulated results are shown in \fig{plot-wilson}. For the finite differencing, we perform simulations with $(\epsilon_1, \epsilon_2) = (0.1,0.0), (0.0,0.1), (0.1,0.1)$, and the same for the parameters $\tilde \epsilon_1,\tilde\epsilon_2$. As before, a Trotterization time step of $\Delta t = 0.2$ is used. While smaller $\epsilon$ is preferred (smaller deformations improve the finite differencing calculation), there is a tradeoff because the resulting smaller finite difference requires a larger number of samples to be collected, scaling like $\epsilon^{-2}$.

In order to test our technique with matter fields, we simulated staggered fermions transforming in a $\mathbb Z_2$ gauge field, in $1+1$ dimensional spacetime. A similar model was also simulated on a quantum computer in \cite{Klco:2018kyo}. The lattice Hamiltonian for this model is
\begin{align}
H = &\sum_{i=1}^{L-1} \left[\frac 1 2 \sigma^{(x)}_{i(i+1)} + \frac {(-1)^i} 2 \bar\chi_{i+1}\sigma^{(z)}_{i(i+1)}\chi_i\right]\nonumber\\
&+
m \sum_{i=1}^L (-1)^i \bar\chi_i \chi_i
\end{align}
where the index $i$ denote lattice sites, the sum is taken over all pairs of adjacent sites, and $\chi$ is a one-component fermionic operator.

Using the Jordan-Wigner procedure, we translate from the fermionic operators $\chi_i$ into a set of bosonic operators $\sigma_i$: $\chi_i = \sigma^{(z)}_1 \cdots \sigma^{(z)}_{i-1} \sigma_i$, where $\sigma_i$ is the lowering operator for site $i$. In one-dimension without periodic boundary conditions, this results in a local bosonic Hamiltonian.
\begin{align}\label{eq:z2ham}
H = \sum_{i=1}^L \bigg[
-&\frac m 2 (-1)^i \sigma^{(z)}_i
+
\sigma^{(x)}_{i (i+1)}
\nonumber\\&+
\frac 1 4 (-1)^i \sigma^{(z)}_{i(i+1)} \left(
\sigma^{(x)}_i
\sigma^{(x)}_{i+1}
+
\sigma^{(y)}_i
\sigma^{(y)}_{i+1}
\right)
\bigg]
\end{align}
There are two distinct families of bosonic operators: the $\sigma_i$ which live at a site and correspond to fermions, and the $\sigma_{i(i+1)}$ on the link from site $i$ to site $i+1$, corresponding to the $\mathbb Z_2$ gauge field. The Trotterization of this Hamiltonian takes four steps, corresponding exactly to the four terms in Eq.~(\ref{eq:z2ham}). In this way full translational invariance (neglecting the boundary) is maintained even in the approximated evolution.

\begin{figure}
\includegraphics[width=\linewidth]{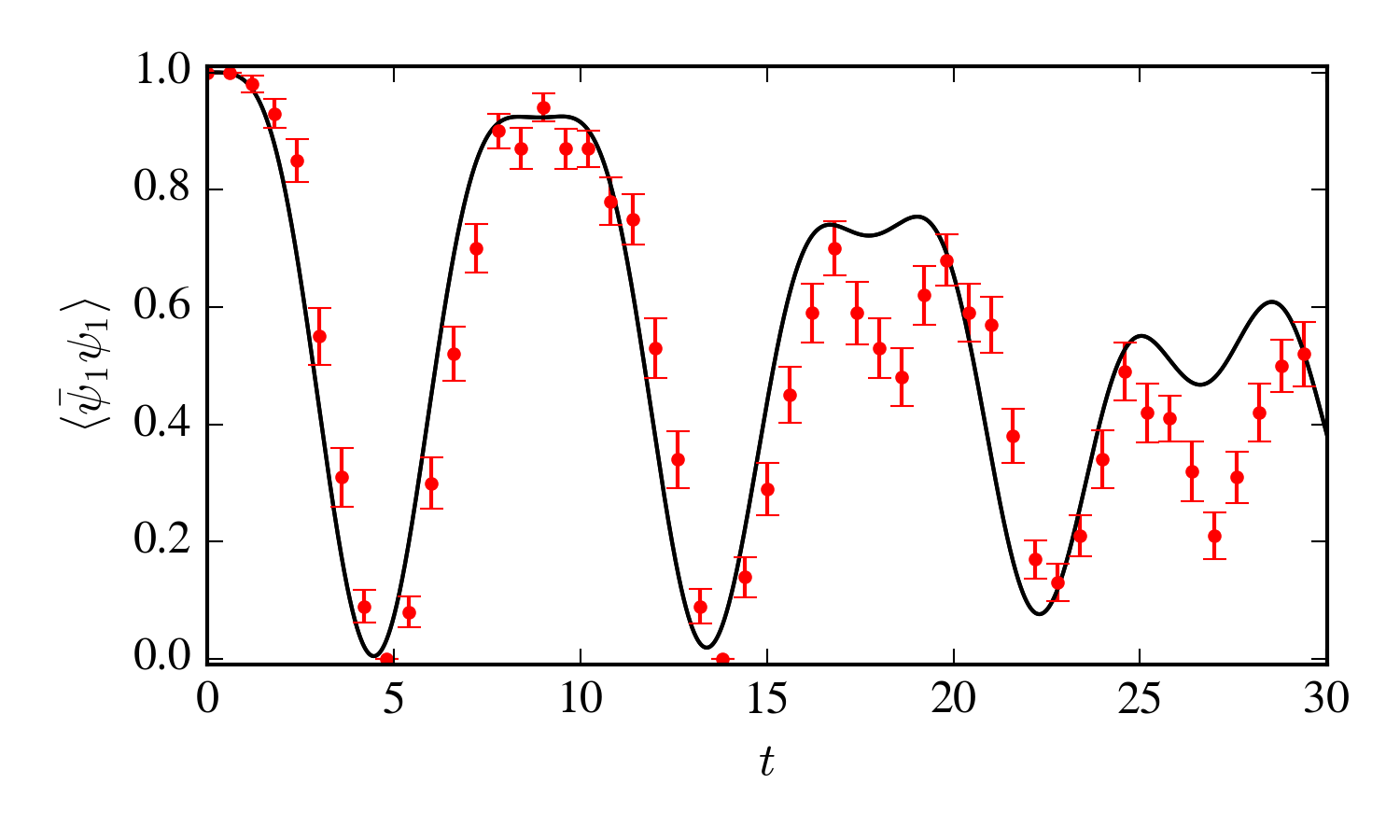}
\caption{Simulation of $\mathbb Z_2$ gauge theory with staggered fermions on four sites.  The expectation value of the number density at the left-most site as a function of $t$. A simulated quantum calculation is shown in red. Differences with the exact result are due to sampling error ($100$ samples per data point) and Trotterization. \label{fig:z2-results}}
\end{figure}

The starting state of this $\mathbb Z_2$ simulation is obtained by projecting the $\left|0000000\right>$ state to the gauge-invariant space, and then applying the gauge-invariant excitation operator $\chi_1^\dagger \chi_2^\dagger \sigma^{(z)}_{12}$.
The number density of the left-most site as a function of $t$ is shown in \fig{z2-results} where the simulated results are compared to the exact values. Since we do not begin in an eigenstate of the Hamiltonian, this is not constant. The difference between the two is due to the Trotterization, and as with the $D_4$ simulation may be made systematically smaller by reducing the lattice spacing. This simulation uses $8$ qubits: $4$ fermion sites, $3$ gauge links, and a single ancillary. For this lattice, $36$ CNOT gates are used per timestep, putting the simulation (slightly) out of reach of present-day processors. A timestep of $\Delta t = 0.6$ is used for $T=30$.

\section{Discussion}\label{sec:discussion}
In this work, we have given a general procedure to construct a unitary time-evolution operator and timelike separated correlation functions for use on an idealized universal quantum computer.  This construction ensures that the Trotterization never mixes the physical and unphysical states within the Hilbert space, provided that the requirements of Sec.~\ref{sec:prerequisites} are implemented exactly.  Further, we show how gauge-invariant correlators (including a temporal Wilson loop) are constructed. We have demonstrated the efficacy of the procedure by the implementation of two gauge theories on a simulated quantum computer. Additionally, we have discussed the \emph{classical} procedure for determining the bare couplings to be used in a \emph{quantum} simulation.

Critically, we have defined a set of primitive constructs which may be used in the construction of gauge theory simulation. The practical and efficient implementation of such registers and gates is an important open question in the literature.
Finally, future work must address the issue of how the behavior of different noisy gates affects the mixing of the physical and unphysical state, whether this can be mitigated by special constructions or error-mitigating codes within our construction, and indeed whether such mitigation is necessary.

\begin{acknowledgments}
H.L., S.L., and Y.Y. are supported by the U.S. Department of Energy under Contract No.~DE-FG02-93ER-40762. The authors thank P. Bedaque and A. Alexandru for conversations on the Hamiltonian formulation of gauge theories, Z. Davoudi for a quantum computing seminar, J. Stryker for helpful comments on this paper.  Finally, the authors would like to thank C. Bonati, M. Cardinali, G. Clemente, L. Cosmai, M. D'Elia, A. Gabbana, S. F. Schifano, R. Tripiccione, and D. Vadacchino for pointing out errors in our manuscript.
\end{acknowledgments}

\appendix

\section{Implementation of a $D_4$-register}\label{app:d4-register}

The group $D_4$ of isometries of the square is treated as a subgroup of $U(2)$ generated by the matrices
\begin{equation}
\left(\begin{matrix}
i & 0\\
0 & -i\\
\end{matrix}\right)
\text{ and }
\left(\begin{matrix}
0 & 1\\
1 & 0\\
\end{matrix}\right)
\text.
\end{equation}
The first matrix represents a $\frac \pi 2$ rotation, and the second represents reflection.

The $D_4$ register is implemented with three qubits. The state $\left|a b c\right>$ corresponds to the matrix
\begin{equation}
\left[\left(\begin{matrix}
0 & 1\\
1 & 0\\
\end{matrix}\right)\right]^a
\left[
\left(\begin{matrix}
i & 0\\
0 & -i\\
\end{matrix}\right)\right]^{2b + c}
\text.
\end{equation}
Thus, the two least-significant bits specify an amount of rotation, to be followed by a flip if the most-significant bit is $1$. This establishes the isomorphism needed between $\mathbb C D_4$ and the three-qubit Hilbert space on a quantum computer.

\begin{figure}
\includegraphics{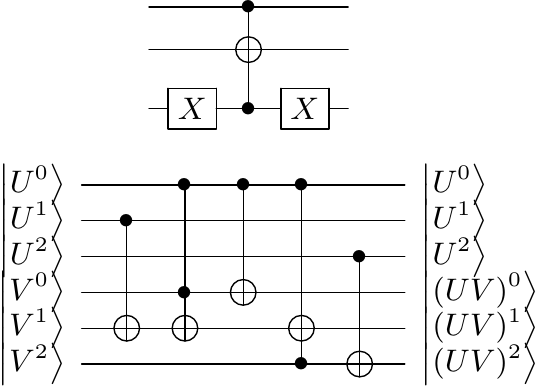}
\caption{Classical-inspired quantum circuits for $D_4$ inversion (top) and multiplication (bottom). The least-significant qubits are shown on the top. For the multiplication circuit, the target register is in the bottom three qubits.\label{fig:classical-circuits}}
\end{figure}

It remains to describe the inversion, multiplication, trace, and Fourier transform circuits. The inversion and multiplication circuits are both equivalent to classical circuits (as they consist exclusively of controlled-not gates), and are shown in \fig{classical-circuits}. The trace circuit is a three-qubit controlled phase gate, defined by $\mathfrak U_{\Tr}(\theta) \left|000\right> = e^{-2 i \theta}$ and $\mathfrak U_{\Tr}(\theta) \left|010\right> = e^{2 i \theta}$ (with all other states picking up no phase).

Finally, the Quantum Fourier Transform which diagonalizes the $D_4$ momentum operator is given by 
\begin{equation}
F = \left(
\begin{matrix}
\frac{1}{\sqrt 8} & \frac{1}{\sqrt 8} & \frac{1}{\sqrt 8} & \frac{1}{\sqrt 8} & \frac{1}{\sqrt 8} & \frac{1}{\sqrt 8} & \frac{1}{\sqrt 8} & \frac{1}{\sqrt 8}\\
\frac{1}{\sqrt 8} & \frac{1}{\sqrt 8} & \frac{1}{\sqrt 8} & \frac{1}{\sqrt 8} & \frac{-1}{\sqrt 8} & \frac{-1}{\sqrt 8} & \frac{-1}{\sqrt 8} & \frac{-1}{\sqrt 8}\\
\frac{1}{\sqrt 8} &-\frac{1}{\sqrt 8} & \frac{1}{\sqrt 8} &-\frac{1}{\sqrt 8} & \frac{1}{\sqrt 8} &-\frac{1}{\sqrt 8} & \frac{1}{\sqrt 8} &-\frac{1}{\sqrt 8}\\
\frac{1}{\sqrt 8} &-\frac{1}{\sqrt 8} & \frac{1}{\sqrt 8} &-\frac{1}{\sqrt 8} &-\frac{1}{\sqrt 8} & \frac{1}{\sqrt 8} &-\frac{1}{\sqrt 8} & \frac{1}{\sqrt 8}\\
\frac 1 2 & 0 & -\frac 1 2 & 0 & \frac 1 2 & 0 & -\frac 1 2 & 0\\
0 & -\frac 1 2 & 0 & \frac 1 2 & 0 & \frac 1 2 & 0 & -\frac 1 2\\
0 & \frac 1 2 & 0 & -\frac 1 2 & 0 & \frac 1 2 & 0 & -\frac 1 2\\
\frac 1 2 & 0 & -\frac 1 2 & 0 & -\frac 1 2 & 0 & \frac 1 2 & 0\\
\end{matrix}
\right)\text.
\end{equation}

\begin{figure}[h]
\includegraphics{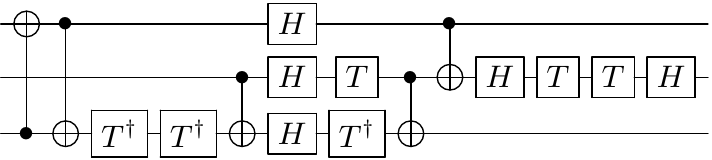}
\caption{Quantum circuit implementing the Fourier transform on a $D_4$ register. The least-significant qubits are on the top. Here, $H$ and $T$ are the Hadamard gate and $\frac\pi 8$ gate, respectively.\label{fig:d4-fourier}}
\end{figure}

Rather than attempt to determine a quantum circuit by hand (although general methods do exist~\cite{2003quant.ph..4064M,1998quant.ph..7064P,beals1997quantum}), we perform a \emph{classical} simulated annealing search to find a short circuit that implements exactly this unitary within the \texttt{qiskit} gate set using Hadamard, \texttt{CNOT}, and $\frac \pi 8$-gates ($T$). The circuit found is shown in \fig{d4-fourier}. The annealing algorithm will be detailed in a future work. Since the size of $F$ does not scale with volume, this method does not suffer from an exponentially large physical Hilbert space.

After being diagonalized by the Fourier transform, the kinetic gate corresponds simply to a phase gate on the most-significant qubit, along with a phase gate on the state $\left|000\right>$:
\begin{equation}
F \mathcal U_K(\theta) F^\dagger \left|a b c\right>
= e^{i \theta_1 \chi_{a =b=c=0}}
e^{i \theta_2 a} \left|a b c\right>
\text.
\end{equation}
The constants $\theta_1$ and $\theta_2$ are coupling-dependent. For $\beta=0.8$, the coupling used in this paper, we have $\theta_1 \approx 1.263$ and $\theta_2 \approx 0.409$.

\section{Positive-Definiteness of the Transfer Matrix}\label{app:positive-definite}

Here we provide a proof that the kinetic part $T_K$ of the transfer matrix for a discrete gauge theory is positive-definite to define a Hamiltonian, a fact used in Sec.~\ref{sec:path-integral}. Unlike prior proofs, we are concerned with positive-definiteness on $\mathcal H$, not just $\mathcal H_P$.

$T_K$ is the tensor product of single-link operators
\begin{equation}
T_K = \bigotimes_{\left<ij\right>}T_K^{(1)}(i,j)
\text,
\end{equation}
and is therefore positive-definite as long as $T_K^{(1)}$ is. The single-link transfer matrix is defined by
\begin{equation}
\left<\tilde g\right|T_K^{(1)}\left|g\right> =
e^{\beta\Re \Tr \left[\rho^\dag\left(\tilde{g}\right) \rho\left(g\right)\right]}
\text,
\end{equation}
where $g, \tilde{g}$ are the elements of $G$, $\rho$ is a faithful representation, and $\beta$ is the inverse coupling. Therefore, our proof need only establish that $T^{(1)}_K$ is positive definite.

\begin{widetext}
It is sufficient to show that $\langle \Psi | T_K^{(1)} |\Psi \rangle > 0$ for all $|\Psi\rangle \ne 0$ in $\mathbb C G$. We must establish that for any function $f(g)$,
\begin{align}
\sum_{g,\tilde{g} \in G} &f^*(\tilde{g})f(g) \langle \tilde{g} |T_K^{(1)}| g \rangle =\sum_{g,\tilde{g} \in G} f^*(\tilde{g})f(g) e^{\beta \Re\Tr[\rho^\dag\left(\tilde{g}\right) \rho\left(g\right)]}
 > 0\text.
\end{align}
We work order-by-order in $\beta$ to see that each term is bounded below by $0$:
 \begin{eqnarray}
\sum_{g,\tilde{g} \in G} f^*(\tilde{g})f(g) e^{\beta \Re\Tr[\rho^\dag\left(\tilde{g}\right) \rho\left(g\right)]}
 = \sum_{n=0}^{\infty} \sum_{i=0}^{n} \frac{\beta^n}{2^n \rm n!}\, _nC_i \sum_{g,\tilde{g} \in G} f^*(\tilde{g})f(g) \Tr[\rho^\dag\left(\tilde{g}\right) \rho\left(g\right)]^i \Tr[\rho\left(\tilde{g}\right) \rho^\dag\left(g\right)]^{n-i}
\end{eqnarray}

where $_nC_i$ are binomial coefficients. Each term indexed by $n,i$ is non-negative:
\begin{align}
\sum_{g,\tilde{g} \in G} f^*(\tilde{g})f(g) &\Tr[\rho^\dag\left(\tilde{g}\right) \rho\left(g\right)]^i \Tr[\rho\left(\tilde{g}\right) \rho^\dag\left(g\right)]^{n-i}\nonumber\\&=\sum_{g,\tilde{g} \in G} f^*(\tilde{g})f(g) \bigg(\sum_{a,b}\rho^\dag\left(\tilde{g}\right)_{a b} \rho\left(g\right)_{b a}\bigg)^i \bigg(\sum_{c,d}\rho\left(\tilde{g}\right)_{c d} \rho^\dag\left(g\right)_{d c}\bigg)^{n-i}\nonumber\\
&= \sum_{\{a\},\{b\},\{c\},\{d\}} |\sum_{g\in G} f(g)\rho\left(g\right)_{b_1 a_1} \cdots \rho\left(g\right)_{b_i a_i} \rho^* \left(g\right)_{c_1 d_1} \cdots \rho^*\left(g\right)_{c_{n-i} d_{n-i}} |^2 \ge 0
\end{align}
where $\{a(b)\} = (a(b)_1,\cdots, a(b)_i)$ and $\{c(d)\} = (c(d)_1,\cdots ,c(d)_{n-i})$.
It remains only to show that $\langle \Psi|T_K^{(1)}|\Psi \rangle = 0$ is impossible. For the inequality to be saturated, the following must hold for any $n, i$ and any combination of matrix components in $\{a\},\{b\},\{c\},\{d\}$.
\begin{equation}
\sum_{g\in G}f(g)\rho\left(g\right)_{b_1 a_1} \cdots \rho\left(g\right)_{b_i a_i} \rho^* \left(g\right)_{c_1 d_1} \cdots \rho^*\left(g\right)_{c_{n-i} d_{n-i}} = 0\label{eq:pol}
\end{equation} 
As the representation $\rho$ is faithful, the space of polynomials on $G$ forms a complete basis, and thus \eq{pol} establishes that the inner product of $f(g)$ with any vector in $\mathbb C G$ vanishes, implying that $f(g) = 0$.

Therefore, for faithful representations $\rho$ (such as the fundamental representation), $T^{(1)}_K$ is positive-definite.
\end{widetext}


\bibliographystyle{apsrev4-1}
\bibliography{wise}

\begin{thebibliography}{87}%
\makeatletter
\providecommand \@ifxundefined [1]{%
 \@ifx{#1\undefined}
}%
\providecommand \@ifnum [1]{%
 \ifnum #1\expandafter \@firstoftwo
 \else \expandafter \@secondoftwo
 \fi
}%
\providecommand \@ifx [1]{%
 \ifx #1\expandafter \@firstoftwo
 \else \expandafter \@secondoftwo
 \fi
}%
\providecommand \natexlab [1]{#1}%
\providecommand \enquote  [1]{``#1''}%
\providecommand \bibnamefont  [1]{#1}%
\providecommand \bibfnamefont [1]{#1}%
\providecommand \citenamefont [1]{#1}%
\providecommand \href@noop [0]{\@secondoftwo}%
\providecommand \href [0]{\begingroup \@sanitize@url \@href}%
\providecommand \@href[1]{\@@startlink{#1}\@@href}%
\providecommand \@@href[1]{\endgroup#1\@@endlink}%
\providecommand \@sanitize@url [0]{\catcode `\\12\catcode `\$12\catcode
  `\&12\catcode `\#12\catcode `\^12\catcode `\_12\catcode `\%12\relax}%
\providecommand \@@startlink[1]{}%
\providecommand \@@endlink[0]{}%
\providecommand \url  [0]{\begingroup\@sanitize@url \@url }%
\providecommand \@url [1]{\endgroup\@href {#1}{\urlprefix }}%
\providecommand \urlprefix  [0]{URL }%
\providecommand \Eprint [0]{\href }%
\providecommand \doibase [0]{http://dx.doi.org/}%
\providecommand \selectlanguage [0]{\@gobble}%
\providecommand \bibinfo  [0]{\@secondoftwo}%
\providecommand \bibfield  [0]{\@secondoftwo}%
\providecommand \translation [1]{[#1]}%
\providecommand \BibitemOpen [0]{}%
\providecommand \bibitemStop [0]{}%
\providecommand \bibitemNoStop [0]{.\EOS\space}%
\providecommand \EOS [0]{\spacefactor3000\relax}%
\providecommand \BibitemShut  [1]{\csname bibitem#1\endcsname}%
\let\auto@bib@innerbib\@empty
\bibitem [{\citenamefont {Feynman}(1982)}]{Feynman:1981tf}%
  \BibitemOpen
  \bibfield  {author} {\bibinfo {author} {\bibfnamefont {R.~P.}\ \bibnamefont
  {Feynman}},\ }\href {\doibase 10.1007/BF02650179} {\bibfield  {journal}
  {\bibinfo  {journal} {Int. J. Theor. Phys.}\ }\textbf {\bibinfo {volume}
  {21}},\ \bibinfo {pages} {467} (\bibinfo {year} {1982})}\BibitemShut
  {NoStop}%
\bibitem [{\citenamefont {Troyer}\ and\ \citenamefont
  {Wiese}(2005)}]{Troyer:2004ge}%
  \BibitemOpen
  \bibfield  {author} {\bibinfo {author} {\bibfnamefont {M.}~\bibnamefont
  {Troyer}}\ and\ \bibinfo {author} {\bibfnamefont {U.-J.}\ \bibnamefont
  {Wiese}},\ }\href {\doibase 10.1103/PhysRevLett.94.170201} {\bibfield
  {journal} {\bibinfo  {journal} {Phys. Rev. Lett.}\ }\textbf {\bibinfo
  {volume} {94}},\ \bibinfo {pages} {170201} (\bibinfo {year} {2005})},\
  \Eprint {http://arxiv.org/abs/cond-mat/0408370} {arXiv:cond-mat/0408370
  [cond-mat]} \BibitemShut {NoStop}%
\bibitem [{\citenamefont {Alexandru}\ \emph {et~al.}(2018)\citenamefont
  {Alexandru}, \citenamefont {Bedaque}, \citenamefont {Lamm}, \citenamefont
  {Lawrence},\ and\ \citenamefont {Warrington}}]{Alexandru:2018ddf}%
  \BibitemOpen
  \bibfield  {author} {\bibinfo {author} {\bibfnamefont {A.}~\bibnamefont
  {Alexandru}}, \bibinfo {author} {\bibfnamefont {P.~F.}\ \bibnamefont
  {Bedaque}}, \bibinfo {author} {\bibfnamefont {H.}~\bibnamefont {Lamm}},
  \bibinfo {author} {\bibfnamefont {S.}~\bibnamefont {Lawrence}}, \ and\
  \bibinfo {author} {\bibfnamefont {N.~C.}\ \bibnamefont {Warrington}},\ }\href
  {\doibase 10.1103/PhysRevLett.121.191602} {\bibfield  {journal} {\bibinfo
  {journal} {Phys. Rev. Lett.}\ }\textbf {\bibinfo {volume} {121}},\ \bibinfo
  {pages} {191602} (\bibinfo {year} {2018})},\ \Eprint
  {http://arxiv.org/abs/1808.09799} {arXiv:1808.09799 [hep-lat]} \BibitemShut
  {NoStop}%
\bibitem [{\citenamefont {Alexandru}\ \emph {et~al.}(2016)\citenamefont
  {Alexandru}, \citenamefont {Basar}, \citenamefont {Bedaque}, \citenamefont
  {Vartak},\ and\ \citenamefont {Warrington}}]{Alexandru:2016gsd}%
  \BibitemOpen
  \bibfield  {author} {\bibinfo {author} {\bibfnamefont {A.}~\bibnamefont
  {Alexandru}}, \bibinfo {author} {\bibfnamefont {G.}~\bibnamefont {Basar}},
  \bibinfo {author} {\bibfnamefont {P.~F.}\ \bibnamefont {Bedaque}}, \bibinfo
  {author} {\bibfnamefont {S.}~\bibnamefont {Vartak}}, \ and\ \bibinfo {author}
  {\bibfnamefont {N.~C.}\ \bibnamefont {Warrington}},\ }\href {\doibase
  10.1103/PhysRevLett.117.081602} {\bibfield  {journal} {\bibinfo  {journal}
  {Phys. Rev. Lett.}\ }\textbf {\bibinfo {volume} {117}},\ \bibinfo {pages}
  {081602} (\bibinfo {year} {2016})},\ \Eprint
  {http://arxiv.org/abs/1605.08040} {arXiv:1605.08040 [hep-lat]} \BibitemShut
  {NoStop}%
\bibitem [{\citenamefont {Dumitrescu}\ \emph {et~al.}(2018)\citenamefont
  {Dumitrescu}, \citenamefont {McCaskey}, \citenamefont {Hagen}, \citenamefont
  {Jansen}, \citenamefont {Morris}, \citenamefont {Papenbrock}, \citenamefont
  {Pooser}, \citenamefont {Dean},\ and\ \citenamefont
  {Lougovski}}]{Dumitrescu:2018njn}%
  \BibitemOpen
  \bibfield  {author} {\bibinfo {author} {\bibfnamefont {E.~F.}\ \bibnamefont
  {Dumitrescu}}, \bibinfo {author} {\bibfnamefont {A.~J.}\ \bibnamefont
  {McCaskey}}, \bibinfo {author} {\bibfnamefont {G.}~\bibnamefont {Hagen}},
  \bibinfo {author} {\bibfnamefont {G.~R.}\ \bibnamefont {Jansen}}, \bibinfo
  {author} {\bibfnamefont {T.~D.}\ \bibnamefont {Morris}}, \bibinfo {author}
  {\bibfnamefont {T.}~\bibnamefont {Papenbrock}}, \bibinfo {author}
  {\bibfnamefont {R.~C.}\ \bibnamefont {Pooser}}, \bibinfo {author}
  {\bibfnamefont {D.~J.}\ \bibnamefont {Dean}}, \ and\ \bibinfo {author}
  {\bibfnamefont {P.}~\bibnamefont {Lougovski}},\ }\href {\doibase
  10.1103/PhysRevLett.120.210501} {\bibfield  {journal} {\bibinfo  {journal}
  {Phys. Rev. Lett.}\ }\textbf {\bibinfo {volume} {120}},\ \bibinfo {pages}
  {210501} (\bibinfo {year} {2018})},\ \Eprint
  {http://arxiv.org/abs/1801.03897} {arXiv:1801.03897 [quant-ph]} \BibitemShut
  {NoStop}%
\bibitem [{\citenamefont {Roggero}\ and\ \citenamefont
  {Carlson}(2018)}]{Roggero:2018hrn}%
  \BibitemOpen
  \bibfield  {author} {\bibinfo {author} {\bibfnamefont {A.}~\bibnamefont
  {Roggero}}\ and\ \bibinfo {author} {\bibfnamefont {J.}~\bibnamefont
  {Carlson}},\ }\href@noop {} {\  (\bibinfo {year} {2018})},\ \Eprint
  {http://arxiv.org/abs/1804.01505} {arXiv:1804.01505 [quant-ph]} \BibitemShut
  {NoStop}%
\bibitem [{\citenamefont {Lu}\ \emph {et~al.}(2018)\citenamefont {Lu} \emph
  {et~al.}}]{Lu:2018pjk}%
  \BibitemOpen
  \bibfield  {author} {\bibinfo {author} {\bibfnamefont {H.-H.}\ \bibnamefont
  {Lu}} \emph {et~al.},\ }\href@noop {} {\  (\bibinfo {year} {2018})},\ \Eprint
  {http://arxiv.org/abs/1810.03959} {arXiv:1810.03959 [quant-ph]} \BibitemShut
  {NoStop}%
\bibitem [{\citenamefont {Martinez}\ \emph {et~al.}(2016)\citenamefont
  {Martinez} \emph {et~al.}}]{Martinez:2016yna}%
  \BibitemOpen
  \bibfield  {author} {\bibinfo {author} {\bibfnamefont {E.~A.}\ \bibnamefont
  {Martinez}} \emph {et~al.},\ }\href {\doibase 10.1038/nature18318} {\bibfield
   {journal} {\bibinfo  {journal} {Nature}\ }\textbf {\bibinfo {volume}
  {534}},\ \bibinfo {pages} {516} (\bibinfo {year} {2016})},\ \Eprint
  {http://arxiv.org/abs/1605.04570} {arXiv:1605.04570 [quant-ph]} \BibitemShut
  {NoStop}%
\bibitem [{\citenamefont {Kokail}\ \emph {et~al.}(2018)\citenamefont {Kokail}
  \emph {et~al.}}]{Kokail:2018eiw}%
  \BibitemOpen
  \bibfield  {author} {\bibinfo {author} {\bibfnamefont {C.}~\bibnamefont
  {Kokail}} \emph {et~al.},\ }\href@noop {} {\  (\bibinfo {year} {2018})},\
  \Eprint {http://arxiv.org/abs/1810.03421} {arXiv:1810.03421 [quant-ph]}
  \BibitemShut {NoStop}%
\bibitem [{\citenamefont {Klco}\ \emph {et~al.}(2018)\citenamefont {Klco},
  \citenamefont {Dumitrescu}, \citenamefont {McCaskey}, \citenamefont {Morris},
  \citenamefont {Pooser}, \citenamefont {Sanz}, \citenamefont {Solano},
  \citenamefont {Lougovski},\ and\ \citenamefont {Savage}}]{Klco:2018kyo}%
  \BibitemOpen
  \bibfield  {author} {\bibinfo {author} {\bibfnamefont {N.}~\bibnamefont
  {Klco}}, \bibinfo {author} {\bibfnamefont {E.~F.}\ \bibnamefont
  {Dumitrescu}}, \bibinfo {author} {\bibfnamefont {A.~J.}\ \bibnamefont
  {McCaskey}}, \bibinfo {author} {\bibfnamefont {T.~D.}\ \bibnamefont
  {Morris}}, \bibinfo {author} {\bibfnamefont {R.~C.}\ \bibnamefont {Pooser}},
  \bibinfo {author} {\bibfnamefont {M.}~\bibnamefont {Sanz}}, \bibinfo {author}
  {\bibfnamefont {E.}~\bibnamefont {Solano}}, \bibinfo {author} {\bibfnamefont
  {P.}~\bibnamefont {Lougovski}}, \ and\ \bibinfo {author} {\bibfnamefont
  {M.~J.}\ \bibnamefont {Savage}},\ }\href {\doibase
  10.1103/PhysRevA.98.032331} {\bibfield  {journal} {\bibinfo  {journal} {Phys.
  Rev.}\ }\textbf {\bibinfo {volume} {A98}},\ \bibinfo {pages} {032331}
  (\bibinfo {year} {2018})},\ \Eprint {http://arxiv.org/abs/1803.03326}
  {arXiv:1803.03326 [quant-ph]} \BibitemShut {NoStop}%
\bibitem [{\citenamefont {Lamm}\ and\ \citenamefont
  {Lawrence}(2018)}]{Lamm:2018siq}%
  \BibitemOpen
  \bibfield  {author} {\bibinfo {author} {\bibfnamefont {H.}~\bibnamefont
  {Lamm}}\ and\ \bibinfo {author} {\bibfnamefont {S.}~\bibnamefont
  {Lawrence}},\ }\href {\doibase 10.1103/PhysRevLett.121.170501} {\bibfield
  {journal} {\bibinfo  {journal} {Phys. Rev. Lett.}\ }\textbf {\bibinfo
  {volume} {121}},\ \bibinfo {pages} {170501} (\bibinfo {year} {2018})},\
  \Eprint {http://arxiv.org/abs/1806.06649} {arXiv:1806.06649 [quant-ph]}
  \BibitemShut {NoStop}%
\bibitem [{\citenamefont {Macridin}\ \emph {et~al.}(2018)\citenamefont
  {Macridin}, \citenamefont {Spentzouris}, \citenamefont {Amundson},\ and\
  \citenamefont {Harnik}}]{Macridin:2018gdw}%
  \BibitemOpen
  \bibfield  {author} {\bibinfo {author} {\bibfnamefont {A.}~\bibnamefont
  {Macridin}}, \bibinfo {author} {\bibfnamefont {P.}~\bibnamefont
  {Spentzouris}}, \bibinfo {author} {\bibfnamefont {J.}~\bibnamefont
  {Amundson}}, \ and\ \bibinfo {author} {\bibfnamefont {R.}~\bibnamefont
  {Harnik}},\ }\href {\doibase 10.1103/PhysRevLett.121.110504} {\bibfield
  {journal} {\bibinfo  {journal} {Phys. Rev. Lett.}\ }\textbf {\bibinfo
  {volume} {121}},\ \bibinfo {pages} {110504} (\bibinfo {year} {2018})},\
  \Eprint {http://arxiv.org/abs/1802.07347} {arXiv:1802.07347 [quant-ph]}
  \BibitemShut {NoStop}%
\bibitem [{\citenamefont {Jordan}\ and\ \citenamefont
  {Wigner}(1928)}]{Jordan:1928wi}%
  \BibitemOpen
  \bibfield  {author} {\bibinfo {author} {\bibfnamefont {P.}~\bibnamefont
  {Jordan}}\ and\ \bibinfo {author} {\bibfnamefont {E.~P.}\ \bibnamefont
  {Wigner}},\ }\href {\doibase 10.1007/BF01331938} {\bibfield  {journal}
  {\bibinfo  {journal} {Z. Phys.}\ }\textbf {\bibinfo {volume} {47}},\ \bibinfo
  {pages} {631} (\bibinfo {year} {1928})}\BibitemShut {NoStop}%
\bibitem [{\citenamefont {Verstraete}\ and\ \citenamefont
  {Cirac}(2005)}]{Verstraete:2005pn}%
  \BibitemOpen
  \bibfield  {author} {\bibinfo {author} {\bibfnamefont {F.}~\bibnamefont
  {Verstraete}}\ and\ \bibinfo {author} {\bibfnamefont {J.~I.}\ \bibnamefont
  {Cirac}},\ }\href {\doibase 10.1088/1742-5468/2005/09/P09012} {\bibfield
  {journal} {\bibinfo  {journal} {J. Stat. Mech.}\ }\textbf {\bibinfo {volume}
  {0509}},\ \bibinfo {pages} {P09012} (\bibinfo {year} {2005})},\ \Eprint
  {http://arxiv.org/abs/cond-mat/0508353} {arXiv:cond-mat/0508353 [cond-mat]}
  \BibitemShut {NoStop}%
\bibitem [{\citenamefont {Zohar}\ and\ \citenamefont
  {Cirac}(2018)}]{Zohar:2018cwb}%
  \BibitemOpen
  \bibfield  {author} {\bibinfo {author} {\bibfnamefont {E.}~\bibnamefont
  {Zohar}}\ and\ \bibinfo {author} {\bibfnamefont {J.~I.}\ \bibnamefont
  {Cirac}},\ }\href {\doibase 10.1103/PhysRevB.98.075119} {\bibfield  {journal}
  {\bibinfo  {journal} {Phys. Rev.}\ }\textbf {\bibinfo {volume} {B98}},\
  \bibinfo {pages} {075119} (\bibinfo {year} {2018})},\ \Eprint
  {http://arxiv.org/abs/1805.05347} {arXiv:1805.05347 [quant-ph]} \BibitemShut
  {NoStop}%
\bibitem [{\citenamefont {{Whitfield}}\ \emph {et~al.}(2016)\citenamefont
  {{Whitfield}}, \citenamefont {{Havl{\'\i}{\v{c}}ek}},\ and\ \citenamefont
  {{Troyer}}}]{2016PhRvA..94c0301W}%
  \BibitemOpen
  \bibfield  {author} {\bibinfo {author} {\bibfnamefont {J.~D.}\ \bibnamefont
  {{Whitfield}}}, \bibinfo {author} {\bibfnamefont {V.}~\bibnamefont
  {{Havl{\'\i}{\v{c}}ek}}}, \ and\ \bibinfo {author} {\bibfnamefont
  {M.}~\bibnamefont {{Troyer}}},\ }\href {\doibase 10.1103/PhysRevA.94.030301}
  {\bibfield  {journal} {\bibinfo  {journal} {Physical Review A}\ }\textbf
  {\bibinfo {volume} {94}},\ \bibinfo {eid} {030301} (\bibinfo {year}
  {2016})},\ \Eprint {http://arxiv.org/abs/1605.09789} {arXiv:1605.09789
  [quant-ph]} \BibitemShut {NoStop}%
\bibitem [{\citenamefont {Hackett}\ \emph {et~al.}(2018)\citenamefont
  {Hackett}, \citenamefont {Howe}, \citenamefont {Hughes}, \citenamefont {Jay},
  \citenamefont {Neil},\ and\ \citenamefont {Simone}}]{Hackett:2018cel}%
  \BibitemOpen
  \bibfield  {author} {\bibinfo {author} {\bibfnamefont {D.~C.}\ \bibnamefont
  {Hackett}}, \bibinfo {author} {\bibfnamefont {K.}~\bibnamefont {Howe}},
  \bibinfo {author} {\bibfnamefont {C.}~\bibnamefont {Hughes}}, \bibinfo
  {author} {\bibfnamefont {W.}~\bibnamefont {Jay}}, \bibinfo {author}
  {\bibfnamefont {E.~T.}\ \bibnamefont {Neil}}, \ and\ \bibinfo {author}
  {\bibfnamefont {J.~N.}\ \bibnamefont {Simone}},\ }\href@noop {} {\  (\bibinfo
  {year} {2018})},\ \Eprint {http://arxiv.org/abs/1811.03629} {arXiv:1811.03629
  [quant-ph]} \BibitemShut {NoStop}%
\bibitem [{\citenamefont {Alexandru}\ \emph {et~al.}(2019)\citenamefont
  {Alexandru}, \citenamefont {Bedaque}, \citenamefont {Harmalkar},
  \citenamefont {Lamm}, \citenamefont {Lawrence},\ and\ \citenamefont
  {Warrington}}]{Alexandru:2019nsa}%
  \BibitemOpen
  \bibfield  {author} {\bibinfo {author} {\bibfnamefont {A.}~\bibnamefont
  {Alexandru}}, \bibinfo {author} {\bibfnamefont {P.~F.}\ \bibnamefont
  {Bedaque}}, \bibinfo {author} {\bibfnamefont {S.}~\bibnamefont {Harmalkar}},
  \bibinfo {author} {\bibfnamefont {H.}~\bibnamefont {Lamm}}, \bibinfo {author}
  {\bibfnamefont {S.}~\bibnamefont {Lawrence}}, \ and\ \bibinfo {author}
  {\bibfnamefont {N.~C.}\ \bibnamefont {Warrington}} (\bibinfo {collaboration}
  {NuQS}),\ }\href@noop {} {\  (\bibinfo {year} {2019})},\ \Eprint
  {http://arxiv.org/abs/1906.11213} {arXiv:1906.11213 [hep-lat]} \BibitemShut
  {NoStop}%
\bibitem [{\citenamefont {Yeter-Aydeniz}\ \emph {et~al.}(2018)\citenamefont
  {Yeter-Aydeniz}, \citenamefont {Dumitrescu}, \citenamefont {McCaskey},
  \citenamefont {Bennink}, \citenamefont {Pooser},\ and\ \citenamefont
  {Siopsis}}]{Yeter-Aydeniz:2018mix}%
  \BibitemOpen
  \bibfield  {author} {\bibinfo {author} {\bibfnamefont {K.}~\bibnamefont
  {Yeter-Aydeniz}}, \bibinfo {author} {\bibfnamefont {E.~F.}\ \bibnamefont
  {Dumitrescu}}, \bibinfo {author} {\bibfnamefont {A.~J.}\ \bibnamefont
  {McCaskey}}, \bibinfo {author} {\bibfnamefont {R.~S.}\ \bibnamefont
  {Bennink}}, \bibinfo {author} {\bibfnamefont {R.~C.}\ \bibnamefont {Pooser}},
  \ and\ \bibinfo {author} {\bibfnamefont {G.}~\bibnamefont {Siopsis}},\
  }\href@noop {} {\  (\bibinfo {year} {2018})},\ \Eprint
  {http://arxiv.org/abs/1811.12332} {arXiv:1811.12332 [quant-ph]} \BibitemShut
  {NoStop}%
\bibitem [{\citenamefont {Klco}\ and\ \citenamefont
  {Savage}(2018)}]{Klco:2018zqz}%
  \BibitemOpen
  \bibfield  {author} {\bibinfo {author} {\bibfnamefont {N.}~\bibnamefont
  {Klco}}\ and\ \bibinfo {author} {\bibfnamefont {M.~J.}\ \bibnamefont
  {Savage}},\ }\href@noop {} {\  (\bibinfo {year} {2018})},\ \Eprint
  {http://arxiv.org/abs/1808.10378} {arXiv:1808.10378 [quant-ph]} \BibitemShut
  {NoStop}%
\bibitem [{\citenamefont {Bazavov}\ \emph {et~al.}(2015)\citenamefont
  {Bazavov}, \citenamefont {Meurice}, \citenamefont {Tsai}, \citenamefont
  {Unmuth-Yockey},\ and\ \citenamefont {Zhang}}]{Bazavov:2015kka}%
  \BibitemOpen
  \bibfield  {author} {\bibinfo {author} {\bibfnamefont {A.}~\bibnamefont
  {Bazavov}}, \bibinfo {author} {\bibfnamefont {Y.}~\bibnamefont {Meurice}},
  \bibinfo {author} {\bibfnamefont {S.-W.}\ \bibnamefont {Tsai}}, \bibinfo
  {author} {\bibfnamefont {J.}~\bibnamefont {Unmuth-Yockey}}, \ and\ \bibinfo
  {author} {\bibfnamefont {J.}~\bibnamefont {Zhang}},\ }\href {\doibase
  10.1103/PhysRevD.92.076003} {\bibfield  {journal} {\bibinfo  {journal} {Phys.
  Rev.}\ }\textbf {\bibinfo {volume} {D92}},\ \bibinfo {pages} {076003}
  (\bibinfo {year} {2015})},\ \Eprint {http://arxiv.org/abs/1503.08354}
  {arXiv:1503.08354 [hep-lat]} \BibitemShut {NoStop}%
\bibitem [{\citenamefont {Zhang}\ \emph {et~al.}(2018)\citenamefont {Zhang},
  \citenamefont {Unmuth-Yockey}, \citenamefont {Zeiher}, \citenamefont
  {Bazavov}, \citenamefont {Tsai},\ and\ \citenamefont
  {Meurice}}]{Zhang:2018ufj}%
  \BibitemOpen
  \bibfield  {author} {\bibinfo {author} {\bibfnamefont {J.}~\bibnamefont
  {Zhang}}, \bibinfo {author} {\bibfnamefont {J.}~\bibnamefont
  {Unmuth-Yockey}}, \bibinfo {author} {\bibfnamefont {J.}~\bibnamefont
  {Zeiher}}, \bibinfo {author} {\bibfnamefont {A.}~\bibnamefont {Bazavov}},
  \bibinfo {author} {\bibfnamefont {S.~W.}\ \bibnamefont {Tsai}}, \ and\
  \bibinfo {author} {\bibfnamefont {Y.}~\bibnamefont {Meurice}},\ }\href
  {\doibase 10.1103/PhysRevLett.121.223201} {\bibfield  {journal} {\bibinfo
  {journal} {Phys. Rev. Lett.}\ }\textbf {\bibinfo {volume} {121}},\ \bibinfo
  {pages} {223201} (\bibinfo {year} {2018})},\ \Eprint
  {http://arxiv.org/abs/1803.11166} {arXiv:1803.11166 [hep-lat]} \BibitemShut
  {NoStop}%
\bibitem [{\citenamefont {Unmuth-Yockey}(2018)}]{Unmuth-Yockey:2018xak}%
  \BibitemOpen
  \bibfield  {author} {\bibinfo {author} {\bibfnamefont {J.~F.}\ \bibnamefont
  {Unmuth-Yockey}},\ }\href@noop {} {\  (\bibinfo {year} {2018})},\ \Eprint
  {http://arxiv.org/abs/1811.05884} {arXiv:1811.05884 [hep-lat]} \BibitemShut
  {NoStop}%
\bibitem [{\citenamefont {Unmuth-Yockey}\ \emph {et~al.}(2018)\citenamefont
  {Unmuth-Yockey}, \citenamefont {Zhang}, \citenamefont {Bazavov},
  \citenamefont {Meurice},\ and\ \citenamefont {Tsai}}]{Unmuth-Yockey:2018ugm}%
  \BibitemOpen
  \bibfield  {author} {\bibinfo {author} {\bibfnamefont {J.}~\bibnamefont
  {Unmuth-Yockey}}, \bibinfo {author} {\bibfnamefont {J.}~\bibnamefont
  {Zhang}}, \bibinfo {author} {\bibfnamefont {A.}~\bibnamefont {Bazavov}},
  \bibinfo {author} {\bibfnamefont {Y.}~\bibnamefont {Meurice}}, \ and\
  \bibinfo {author} {\bibfnamefont {S.-W.}\ \bibnamefont {Tsai}},\ }\href
  {\doibase 10.1103/PhysRevD.98.094511} {\bibfield  {journal} {\bibinfo
  {journal} {Phys. Rev.}\ }\textbf {\bibinfo {volume} {D98}},\ \bibinfo {pages}
  {094511} (\bibinfo {year} {2018})},\ \Eprint
  {http://arxiv.org/abs/1807.09186} {arXiv:1807.09186 [hep-lat]} \BibitemShut
  {NoStop}%
\bibitem [{\citenamefont {Kaplan}\ and\ \citenamefont
  {Stryker}(2018)}]{Kaplan:2018vnj}%
  \BibitemOpen
  \bibfield  {author} {\bibinfo {author} {\bibfnamefont {D.~B.}\ \bibnamefont
  {Kaplan}}\ and\ \bibinfo {author} {\bibfnamefont {J.~R.}\ \bibnamefont
  {Stryker}},\ }\href@noop {} {\  (\bibinfo {year} {2018})},\ \Eprint
  {http://arxiv.org/abs/1806.08797} {arXiv:1806.08797 [hep-lat]} \BibitemShut
  {NoStop}%
\bibitem [{\citenamefont {Meurice}(2019)}]{Meurice:2019ddf}%
  \BibitemOpen
  \bibfield  {author} {\bibinfo {author} {\bibfnamefont {Y.}~\bibnamefont
  {Meurice}},\ }\href@noop {} {\  (\bibinfo {year} {2019})},\ \Eprint
  {http://arxiv.org/abs/1903.01918} {arXiv:1903.01918 [hep-lat]} \BibitemShut
  {NoStop}%
\bibitem [{\citenamefont {Zache}\ \emph {et~al.}(2018)\citenamefont {Zache},
  \citenamefont {Hebenstreit}, \citenamefont {Jendrzejewski}, \citenamefont
  {Oberthaler}, \citenamefont {Berges},\ and\ \citenamefont
  {Hauke}}]{Zache:2018jbt}%
  \BibitemOpen
  \bibfield  {author} {\bibinfo {author} {\bibfnamefont {T.~V.}\ \bibnamefont
  {Zache}}, \bibinfo {author} {\bibfnamefont {F.}~\bibnamefont {Hebenstreit}},
  \bibinfo {author} {\bibfnamefont {F.}~\bibnamefont {Jendrzejewski}}, \bibinfo
  {author} {\bibfnamefont {M.~K.}\ \bibnamefont {Oberthaler}}, \bibinfo
  {author} {\bibfnamefont {J.}~\bibnamefont {Berges}}, \ and\ \bibinfo {author}
  {\bibfnamefont {P.}~\bibnamefont {Hauke}},\ }\href {\doibase
  10.1088/2058-9565/aac33b} {\bibfield  {journal} {\bibinfo  {journal} {Sci.
  Technol.}\ }\textbf {\bibinfo {volume} {3}},\ \bibinfo {pages} {034010}
  (\bibinfo {year} {2018})},\ \Eprint {http://arxiv.org/abs/1802.06704}
  {arXiv:1802.06704 [cond-mat.quant-gas]} \BibitemShut {NoStop}%
\bibitem [{\citenamefont {Raychowdhury}\ and\ \citenamefont
  {Stryker}(2018)}]{Raychowdhury:2018osk}%
  \BibitemOpen
  \bibfield  {author} {\bibinfo {author} {\bibfnamefont {I.}~\bibnamefont
  {Raychowdhury}}\ and\ \bibinfo {author} {\bibfnamefont {J.~R.}\ \bibnamefont
  {Stryker}},\ }\href@noop {} {\  (\bibinfo {year} {2018})},\ \Eprint
  {http://arxiv.org/abs/1812.07554} {arXiv:1812.07554 [hep-lat]} \BibitemShut
  {NoStop}%
\bibitem [{\citenamefont {Peruzzo}\ \emph {et~al.}(2014)\citenamefont
  {Peruzzo}, \citenamefont {McClean}, \citenamefont {Shadbolt}, \citenamefont
  {Yung}, \citenamefont {Zhou}, \citenamefont {Love}, \citenamefont
  {Aspuru-Guzik},\ and\ \citenamefont {O’brien}}]{peruzzo2014variational}%
  \BibitemOpen
  \bibfield  {author} {\bibinfo {author} {\bibfnamefont {A.}~\bibnamefont
  {Peruzzo}}, \bibinfo {author} {\bibfnamefont {J.}~\bibnamefont {McClean}},
  \bibinfo {author} {\bibfnamefont {P.}~\bibnamefont {Shadbolt}}, \bibinfo
  {author} {\bibfnamefont {M.-H.}\ \bibnamefont {Yung}}, \bibinfo {author}
  {\bibfnamefont {X.-Q.}\ \bibnamefont {Zhou}}, \bibinfo {author}
  {\bibfnamefont {P.~J.}\ \bibnamefont {Love}}, \bibinfo {author}
  {\bibfnamefont {A.}~\bibnamefont {Aspuru-Guzik}}, \ and\ \bibinfo {author}
  {\bibfnamefont {J.~L.}\ \bibnamefont {O’brien}},\ }\href@noop {} {\bibfield
   {journal} {\bibinfo  {journal} {Nature communications}\ }\textbf {\bibinfo
  {volume} {5}},\ \bibinfo {pages} {4213} (\bibinfo {year} {2014})}\BibitemShut
  {NoStop}%
\bibitem [{\citenamefont {Abrams}\ and\ \citenamefont
  {Lloyd}(1999)}]{Abrams:1998pd}%
  \BibitemOpen
  \bibfield  {author} {\bibinfo {author} {\bibfnamefont {D.~S.}\ \bibnamefont
  {Abrams}}\ and\ \bibinfo {author} {\bibfnamefont {S.}~\bibnamefont {Lloyd}},\
  }\href {\doibase 10.1103/PhysRevLett.83.5162} {\bibfield  {journal} {\bibinfo
   {journal} {Phys. Rev. Lett.}\ }\textbf {\bibinfo {volume} {83}},\ \bibinfo
  {pages} {5162} (\bibinfo {year} {1999})},\ \Eprint
  {http://arxiv.org/abs/quant-ph/9807070} {arXiv:quant-ph/9807070 [quant-ph]}
  \BibitemShut {NoStop}%
\bibitem [{\citenamefont {Nielsen}\ and\ \citenamefont
  {Chuang}(2000)}]{nielsen2000quantum}%
  \BibitemOpen
  \bibfield  {author} {\bibinfo {author} {\bibfnamefont {M.~A.}\ \bibnamefont
  {Nielsen}}\ and\ \bibinfo {author} {\bibfnamefont {I.~L.}\ \bibnamefont
  {Chuang}},\ }\href@noop {} {\enquote {\bibinfo {title} {Quantum computation
  and quantum information},}\ } (\bibinfo {year} {2000})\BibitemShut {NoStop}%
\bibitem [{\citenamefont {Wiebe}\ and\ \citenamefont
  {Granade}(2016)}]{PhysRevLett.117.010503}%
  \BibitemOpen
  \bibfield  {author} {\bibinfo {author} {\bibfnamefont {N.}~\bibnamefont
  {Wiebe}}\ and\ \bibinfo {author} {\bibfnamefont {C.}~\bibnamefont
  {Granade}},\ }\href {\doibase 10.1103/PhysRevLett.117.010503} {\bibfield
  {journal} {\bibinfo  {journal} {Phys. Rev. Lett.}\ }\textbf {\bibinfo
  {volume} {117}},\ \bibinfo {pages} {010503} (\bibinfo {year}
  {2016})}\BibitemShut {NoStop}%
\bibitem [{\citenamefont {Farhi}\ \emph {et~al.}(2000)\citenamefont {Farhi},
  \citenamefont {Goldstone}, \citenamefont {Gutmann},\ and\ \citenamefont
  {Sipser}}]{farhi2000quantum}%
  \BibitemOpen
  \bibfield  {author} {\bibinfo {author} {\bibfnamefont {E.}~\bibnamefont
  {Farhi}}, \bibinfo {author} {\bibfnamefont {J.}~\bibnamefont {Goldstone}},
  \bibinfo {author} {\bibfnamefont {S.}~\bibnamefont {Gutmann}}, \ and\
  \bibinfo {author} {\bibfnamefont {M.}~\bibnamefont {Sipser}},\ }\href@noop {}
  {\bibfield  {journal} {\bibinfo  {journal} {arXiv:0001106}\ } (\bibinfo
  {year} {2000})}\BibitemShut {NoStop}%
\bibitem [{\citenamefont {Farhi}\ \emph {et~al.}(2001)\citenamefont {Farhi},
  \citenamefont {Goldstone}, \citenamefont {Gutmann}, \citenamefont {Lapan},
  \citenamefont {Lundgren},\ and\ \citenamefont {Preda}}]{Farhi472}%
  \BibitemOpen
  \bibfield  {author} {\bibinfo {author} {\bibfnamefont {E.}~\bibnamefont
  {Farhi}}, \bibinfo {author} {\bibfnamefont {J.}~\bibnamefont {Goldstone}},
  \bibinfo {author} {\bibfnamefont {S.}~\bibnamefont {Gutmann}}, \bibinfo
  {author} {\bibfnamefont {J.}~\bibnamefont {Lapan}}, \bibinfo {author}
  {\bibfnamefont {A.}~\bibnamefont {Lundgren}}, \ and\ \bibinfo {author}
  {\bibfnamefont {D.}~\bibnamefont {Preda}},\ }\href {\doibase
  10.1126/science.1057726} {\bibfield  {journal} {\bibinfo  {journal}
  {Science}\ }\textbf {\bibinfo {volume} {292}},\ \bibinfo {pages} {472}
  (\bibinfo {year} {2001})}\BibitemShut {NoStop}%
\bibitem [{\citenamefont {Kaplan}\ \emph {et~al.}(2017)\citenamefont {Kaplan},
  \citenamefont {Klco},\ and\ \citenamefont {Roggero}}]{Kaplan:2017ccd}%
  \BibitemOpen
  \bibfield  {author} {\bibinfo {author} {\bibfnamefont {D.~B.}\ \bibnamefont
  {Kaplan}}, \bibinfo {author} {\bibfnamefont {N.}~\bibnamefont {Klco}}, \ and\
  \bibinfo {author} {\bibfnamefont {A.}~\bibnamefont {Roggero}},\ }\href@noop
  {} {\  (\bibinfo {year} {2017})},\ \Eprint {http://arxiv.org/abs/1709.08250}
  {arXiv:1709.08250 [quant-ph]} \BibitemShut {NoStop}%
\bibitem [{\citenamefont {{Bilgin}}\ and\ \citenamefont
  {{Boixo}}(2010)}]{2010PhRvL.105q0405B}%
  \BibitemOpen
  \bibfield  {author} {\bibinfo {author} {\bibfnamefont {E.}~\bibnamefont
  {{Bilgin}}}\ and\ \bibinfo {author} {\bibfnamefont {S.}~\bibnamefont
  {{Boixo}}},\ }\href {\doibase 10.1103/PhysRevLett.105.170405} {\bibfield
  {journal} {\bibinfo  {journal} {Physical Review Letters}\ }\textbf {\bibinfo
  {volume} {105}},\ \bibinfo {eid} {170405} (\bibinfo {year} {2010})},\ \Eprint
  {http://arxiv.org/abs/1008.4162} {arXiv:1008.4162 [quant-ph]} \BibitemShut
  {NoStop}%
\bibitem [{\citenamefont {Jordan}\ \emph {et~al.}(2012)\citenamefont {Jordan},
  \citenamefont {Lee},\ and\ \citenamefont {Preskill}}]{Jordan:2011ne}%
  \BibitemOpen
  \bibfield  {author} {\bibinfo {author} {\bibfnamefont {S.~P.}\ \bibnamefont
  {Jordan}}, \bibinfo {author} {\bibfnamefont {K.~S.~M.}\ \bibnamefont {Lee}},
  \ and\ \bibinfo {author} {\bibfnamefont {J.}~\bibnamefont {Preskill}},\
  }\href {\doibase 10.1126/science.1217069} {\bibfield  {journal} {\bibinfo
  {journal} {Science}\ }\textbf {\bibinfo {volume} {336}},\ \bibinfo {pages}
  {1130} (\bibinfo {year} {2012})},\ \Eprint {http://arxiv.org/abs/1111.3633}
  {arXiv:1111.3633 [quant-ph]} \BibitemShut {NoStop}%
\bibitem [{\citenamefont {Jordan}\ \emph {et~al.}(2011)\citenamefont {Jordan},
  \citenamefont {Lee},\ and\ \citenamefont {Preskill}}]{Jordan:2011ci}%
  \BibitemOpen
  \bibfield  {author} {\bibinfo {author} {\bibfnamefont {S.~P.}\ \bibnamefont
  {Jordan}}, \bibinfo {author} {\bibfnamefont {K.~S.~M.}\ \bibnamefont {Lee}},
  \ and\ \bibinfo {author} {\bibfnamefont {J.}~\bibnamefont {Preskill}},\
  }\href@noop {} {\  (\bibinfo {year} {2011})},\ \bibinfo {note} {[Quant. Inf.
  Comput.14,1014(2014)]},\ \Eprint {http://arxiv.org/abs/1112.4833}
  {arXiv:1112.4833 [hep-th]} \BibitemShut {NoStop}%
\bibitem [{\citenamefont {García-Álvarez}\ \emph {et~al.}(2015)\citenamefont
  {García-Álvarez}, \citenamefont {Casanova}, \citenamefont {Mezzacapo},
  \citenamefont {Egusquiza}, \citenamefont {Lamata}, \citenamefont {Romero},\
  and\ \citenamefont {Solano}}]{Garcia-Alvarez:2014uda}%
  \BibitemOpen
  \bibfield  {author} {\bibinfo {author} {\bibfnamefont {L.}~\bibnamefont
  {García-Álvarez}}, \bibinfo {author} {\bibfnamefont {J.}~\bibnamefont
  {Casanova}}, \bibinfo {author} {\bibfnamefont {A.}~\bibnamefont {Mezzacapo}},
  \bibinfo {author} {\bibfnamefont {I.~L.}\ \bibnamefont {Egusquiza}}, \bibinfo
  {author} {\bibfnamefont {L.}~\bibnamefont {Lamata}}, \bibinfo {author}
  {\bibfnamefont {G.}~\bibnamefont {Romero}}, \ and\ \bibinfo {author}
  {\bibfnamefont {E.}~\bibnamefont {Solano}},\ }\href {\doibase
  10.1103/PhysRevLett.114.070502} {\bibfield  {journal} {\bibinfo  {journal}
  {Phys. Rev. Lett.}\ }\textbf {\bibinfo {volume} {114}},\ \bibinfo {pages}
  {070502} (\bibinfo {year} {2015})},\ \Eprint {http://arxiv.org/abs/1404.2868}
  {arXiv:1404.2868 [quant-ph]} \BibitemShut {NoStop}%
\bibitem [{\citenamefont {Jordan}\ \emph {et~al.}(2014)\citenamefont {Jordan},
  \citenamefont {Lee},\ and\ \citenamefont {Preskill}}]{Jordan:2014tma}%
  \BibitemOpen
  \bibfield  {author} {\bibinfo {author} {\bibfnamefont {S.~P.}\ \bibnamefont
  {Jordan}}, \bibinfo {author} {\bibfnamefont {K.~S.~M.}\ \bibnamefont {Lee}},
  \ and\ \bibinfo {author} {\bibfnamefont {J.}~\bibnamefont {Preskill}},\
  }\href@noop {} {\  (\bibinfo {year} {2014})},\ \Eprint
  {http://arxiv.org/abs/1404.7115} {arXiv:1404.7115 [hep-th]} \BibitemShut
  {NoStop}%
\bibitem [{\citenamefont {Jordan}\ \emph {et~al.}(2017)\citenamefont {Jordan},
  \citenamefont {Krovi}, \citenamefont {Lee},\ and\ \citenamefont
  {Preskill}}]{Jordan:2017lea}%
  \BibitemOpen
  \bibfield  {author} {\bibinfo {author} {\bibfnamefont {S.~P.}\ \bibnamefont
  {Jordan}}, \bibinfo {author} {\bibfnamefont {H.}~\bibnamefont {Krovi}},
  \bibinfo {author} {\bibfnamefont {K.~S.~M.}\ \bibnamefont {Lee}}, \ and\
  \bibinfo {author} {\bibfnamefont {J.}~\bibnamefont {Preskill}},\ }\href@noop
  {} {\  (\bibinfo {year} {2017})},\ \Eprint {http://arxiv.org/abs/1703.00454}
  {arXiv:1703.00454 [quant-ph]} \BibitemShut {NoStop}%
\bibitem [{\citenamefont {Hamed~Moosavian}\ and\ \citenamefont
  {Jordan}(2018)}]{Moosavian:2017tkv}%
  \BibitemOpen
  \bibfield  {author} {\bibinfo {author} {\bibfnamefont {A.}~\bibnamefont
  {Hamed~Moosavian}}\ and\ \bibinfo {author} {\bibfnamefont {S.}~\bibnamefont
  {Jordan}},\ }\href {\doibase 10.1103/PhysRevA.98.012332} {\bibfield
  {journal} {\bibinfo  {journal} {Phys. Rev.}\ }\textbf {\bibinfo {volume}
  {A98}},\ \bibinfo {pages} {012332} (\bibinfo {year} {2018})},\ \Eprint
  {http://arxiv.org/abs/1711.04006} {arXiv:1711.04006 [quant-ph]} \BibitemShut
  {NoStop}%
\bibitem [{\citenamefont {Pedernales}\ \emph {et~al.}(2014)\citenamefont
  {Pedernales}, \citenamefont {Di~Candia}, \citenamefont {Egusquiza},
  \citenamefont {Casanova},\ and\ \citenamefont
  {Solano}}]{PhysRevLett.113.020505}%
  \BibitemOpen
  \bibfield  {author} {\bibinfo {author} {\bibfnamefont {J.~S.}\ \bibnamefont
  {Pedernales}}, \bibinfo {author} {\bibfnamefont {R.}~\bibnamefont
  {Di~Candia}}, \bibinfo {author} {\bibfnamefont {I.~L.}\ \bibnamefont
  {Egusquiza}}, \bibinfo {author} {\bibfnamefont {J.}~\bibnamefont {Casanova}},
  \ and\ \bibinfo {author} {\bibfnamefont {E.}~\bibnamefont {Solano}},\ }\href
  {\doibase 10.1103/PhysRevLett.113.020505} {\bibfield  {journal} {\bibinfo
  {journal} {Phys. Rev. Lett.}\ }\textbf {\bibinfo {volume} {113}},\ \bibinfo
  {pages} {020505} (\bibinfo {year} {2014})}\BibitemShut {NoStop}%
\bibitem [{\citenamefont {Bermudez}\ \emph {et~al.}(2017)\citenamefont
  {Bermudez}, \citenamefont {Aarts},\ and\ \citenamefont
  {Müller}}]{Bermudez:2017yrq}%
  \BibitemOpen
  \bibfield  {author} {\bibinfo {author} {\bibfnamefont {A.}~\bibnamefont
  {Bermudez}}, \bibinfo {author} {\bibfnamefont {G.}~\bibnamefont {Aarts}}, \
  and\ \bibinfo {author} {\bibfnamefont {M.}~\bibnamefont {Müller}},\ }\href
  {\doibase 10.1103/PhysRevX.7.041012} {\bibfield  {journal} {\bibinfo
  {journal} {Phys. Rev.}\ }\textbf {\bibinfo {volume} {X7}},\ \bibinfo {pages}
  {041012} (\bibinfo {year} {2017})},\ \Eprint
  {http://arxiv.org/abs/1704.02877} {arXiv:1704.02877 [quant-ph]} \BibitemShut
  {NoStop}%
\bibitem [{\citenamefont {Peardon}\ \emph {et~al.}(2009)\citenamefont
  {Peardon}, \citenamefont {Bulava}, \citenamefont {Foley}, \citenamefont
  {Morningstar}, \citenamefont {Dudek}, \citenamefont {Edwards}, \citenamefont
  {Joo}, \citenamefont {Lin}, \citenamefont {Richards},\ and\ \citenamefont
  {Juge}}]{Peardon:2009gh}%
  \BibitemOpen
  \bibfield  {author} {\bibinfo {author} {\bibfnamefont {M.}~\bibnamefont
  {Peardon}}, \bibinfo {author} {\bibfnamefont {J.}~\bibnamefont {Bulava}},
  \bibinfo {author} {\bibfnamefont {J.}~\bibnamefont {Foley}}, \bibinfo
  {author} {\bibfnamefont {C.}~\bibnamefont {Morningstar}}, \bibinfo {author}
  {\bibfnamefont {J.}~\bibnamefont {Dudek}}, \bibinfo {author} {\bibfnamefont
  {R.~G.}\ \bibnamefont {Edwards}}, \bibinfo {author} {\bibfnamefont
  {B.}~\bibnamefont {Joo}}, \bibinfo {author} {\bibfnamefont {H.-W.}\
  \bibnamefont {Lin}}, \bibinfo {author} {\bibfnamefont {D.~G.}\ \bibnamefont
  {Richards}}, \ and\ \bibinfo {author} {\bibfnamefont {K.~J.}\ \bibnamefont
  {Juge}} (\bibinfo {collaboration} {Hadron Spectrum}),\ }\href {\doibase
  10.1103/PhysRevD.80.054506} {\bibfield  {journal} {\bibinfo  {journal} {Phys.
  Rev.}\ }\textbf {\bibinfo {volume} {D80}},\ \bibinfo {pages} {054506}
  (\bibinfo {year} {2009})},\ \Eprint {http://arxiv.org/abs/0905.2160}
  {arXiv:0905.2160 [hep-lat]} \BibitemShut {NoStop}%
\bibitem [{\citenamefont {Haah}\ \emph {et~al.}(2018)\citenamefont {Haah},
  \citenamefont {Hastings}, \citenamefont {Kothari},\ and\ \citenamefont
  {Low}}]{haah2018quantum}%
  \BibitemOpen
  \bibfield  {author} {\bibinfo {author} {\bibfnamefont {J.}~\bibnamefont
  {Haah}}, \bibinfo {author} {\bibfnamefont {M.~B.}\ \bibnamefont {Hastings}},
  \bibinfo {author} {\bibfnamefont {R.}~\bibnamefont {Kothari}}, \ and\
  \bibinfo {author} {\bibfnamefont {G.~H.}\ \bibnamefont {Low}},\ }\href@noop
  {} {\bibfield  {journal} {\bibinfo  {journal} {arXiv:1801.03922}\ } (\bibinfo
  {year} {2018})}\BibitemShut {NoStop}%
\bibitem [{\citenamefont {Banerjee}\ \emph {et~al.}(2012)\citenamefont
  {Banerjee}, \citenamefont {Dalmonte}, \citenamefont {Muller}, \citenamefont
  {Rico}, \citenamefont {Stebler}, \citenamefont {Wiese},\ and\ \citenamefont
  {Zoller}}]{Banerjee:2012pg}%
  \BibitemOpen
  \bibfield  {author} {\bibinfo {author} {\bibfnamefont {D.}~\bibnamefont
  {Banerjee}}, \bibinfo {author} {\bibfnamefont {M.}~\bibnamefont {Dalmonte}},
  \bibinfo {author} {\bibfnamefont {M.}~\bibnamefont {Muller}}, \bibinfo
  {author} {\bibfnamefont {E.}~\bibnamefont {Rico}}, \bibinfo {author}
  {\bibfnamefont {P.}~\bibnamefont {Stebler}}, \bibinfo {author} {\bibfnamefont
  {U.~J.}\ \bibnamefont {Wiese}}, \ and\ \bibinfo {author} {\bibfnamefont
  {P.}~\bibnamefont {Zoller}},\ }\href {\doibase
  10.1103/PhysRevLett.109.175302} {\bibfield  {journal} {\bibinfo  {journal}
  {Phys. Rev. Lett.}\ }\textbf {\bibinfo {volume} {109}},\ \bibinfo {pages}
  {175302} (\bibinfo {year} {2012})},\ \Eprint {http://arxiv.org/abs/1205.6366}
  {arXiv:1205.6366 [cond-mat.quant-gas]} \BibitemShut {NoStop}%
\bibitem [{\citenamefont {Hauke}\ \emph {et~al.}(2013)\citenamefont {Hauke},
  \citenamefont {Marcos}, \citenamefont {Dalmonte},\ and\ \citenamefont
  {Zoller}}]{Hauke:2013jga}%
  \BibitemOpen
  \bibfield  {author} {\bibinfo {author} {\bibfnamefont {P.}~\bibnamefont
  {Hauke}}, \bibinfo {author} {\bibfnamefont {D.}~\bibnamefont {Marcos}},
  \bibinfo {author} {\bibfnamefont {M.}~\bibnamefont {Dalmonte}}, \ and\
  \bibinfo {author} {\bibfnamefont {P.}~\bibnamefont {Zoller}},\ }\href
  {\doibase 10.1103/PhysRevX.3.041018} {\bibfield  {journal} {\bibinfo
  {journal} {Phys. Rev.}\ }\textbf {\bibinfo {volume} {X3}},\ \bibinfo {pages}
  {041018} (\bibinfo {year} {2013})},\ \Eprint {http://arxiv.org/abs/1306.2162}
  {arXiv:1306.2162 [cond-mat.quant-gas]} \BibitemShut {NoStop}%
\bibitem [{\citenamefont {Kuno}\ \emph {et~al.}(2015)\citenamefont {Kuno},
  \citenamefont {Kasamatsu}, \citenamefont {Takahashi}, \citenamefont
  {Ichinose},\ and\ \citenamefont {Matsui}}]{Kuno:2014npa}%
  \BibitemOpen
  \bibfield  {author} {\bibinfo {author} {\bibfnamefont {Y.}~\bibnamefont
  {Kuno}}, \bibinfo {author} {\bibfnamefont {K.}~\bibnamefont {Kasamatsu}},
  \bibinfo {author} {\bibfnamefont {Y.}~\bibnamefont {Takahashi}}, \bibinfo
  {author} {\bibfnamefont {I.}~\bibnamefont {Ichinose}}, \ and\ \bibinfo
  {author} {\bibfnamefont {T.}~\bibnamefont {Matsui}},\ }\href {\doibase
  10.1088/1367-2630/17/6/063005} {\bibfield  {journal} {\bibinfo  {journal}
  {New J. Phys.}\ }\textbf {\bibinfo {volume} {17}},\ \bibinfo {pages} {063005}
  (\bibinfo {year} {2015})},\ \Eprint {http://arxiv.org/abs/1412.7605}
  {arXiv:1412.7605 [cond-mat.quant-gas]} \BibitemShut {NoStop}%
\bibitem [{\citenamefont {Kuno}\ \emph {et~al.}(2017)\citenamefont {Kuno},
  \citenamefont {Sakane}, \citenamefont {Kasamatsu}, \citenamefont {Ichinose},\
  and\ \citenamefont {Matsui}}]{Kuno:2016ipi}%
  \BibitemOpen
  \bibfield  {author} {\bibinfo {author} {\bibfnamefont {Y.}~\bibnamefont
  {Kuno}}, \bibinfo {author} {\bibfnamefont {S.}~\bibnamefont {Sakane}},
  \bibinfo {author} {\bibfnamefont {K.}~\bibnamefont {Kasamatsu}}, \bibinfo
  {author} {\bibfnamefont {I.}~\bibnamefont {Ichinose}}, \ and\ \bibinfo
  {author} {\bibfnamefont {T.}~\bibnamefont {Matsui}},\ }\href {\doibase
  10.1103/PhysRevD.95.094507} {\bibfield  {journal} {\bibinfo  {journal} {Phys.
  Rev.}\ }\textbf {\bibinfo {volume} {D95}},\ \bibinfo {pages} {094507}
  (\bibinfo {year} {2017})},\ \Eprint {http://arxiv.org/abs/1605.00333}
  {arXiv:1605.00333 [cond-mat.quant-gas]} \BibitemShut {NoStop}%
\bibitem [{\citenamefont {Marcos}\ \emph {et~al.}(2014)\citenamefont {Marcos},
  \citenamefont {Widmer}, \citenamefont {Rico}, \citenamefont {Hafezi},
  \citenamefont {Rabl}, \citenamefont {Wiese},\ and\ \citenamefont
  {Zoller}}]{Marcos:2014lda}%
  \BibitemOpen
  \bibfield  {author} {\bibinfo {author} {\bibfnamefont {D.}~\bibnamefont
  {Marcos}}, \bibinfo {author} {\bibfnamefont {P.}~\bibnamefont {Widmer}},
  \bibinfo {author} {\bibfnamefont {E.}~\bibnamefont {Rico}}, \bibinfo {author}
  {\bibfnamefont {M.}~\bibnamefont {Hafezi}}, \bibinfo {author} {\bibfnamefont
  {P.}~\bibnamefont {Rabl}}, \bibinfo {author} {\bibfnamefont {U.~J.}\
  \bibnamefont {Wiese}}, \ and\ \bibinfo {author} {\bibfnamefont
  {P.}~\bibnamefont {Zoller}},\ }\href {\doibase 10.1016/j.aop.2014.09.011}
  {\bibfield  {journal} {\bibinfo  {journal} {Annals Phys.}\ }\textbf {\bibinfo
  {volume} {351}},\ \bibinfo {pages} {634} (\bibinfo {year} {2014})},\ \Eprint
  {http://arxiv.org/abs/1407.6066} {arXiv:1407.6066 [quant-ph]} \BibitemShut
  {NoStop}%
\bibitem [{\citenamefont {Muschik}\ \emph {et~al.}(2017)\citenamefont
  {Muschik}, \citenamefont {Heyl}, \citenamefont {Martinez}, \citenamefont
  {Monz}, \citenamefont {Schindler}, \citenamefont {Vogell}, \citenamefont
  {Dalmonte}, \citenamefont {Hauke}, \citenamefont {Blatt},\ and\ \citenamefont
  {Zoller}}]{Muschik:2016tws}%
  \BibitemOpen
  \bibfield  {author} {\bibinfo {author} {\bibfnamefont {C.}~\bibnamefont
  {Muschik}}, \bibinfo {author} {\bibfnamefont {M.}~\bibnamefont {Heyl}},
  \bibinfo {author} {\bibfnamefont {E.}~\bibnamefont {Martinez}}, \bibinfo
  {author} {\bibfnamefont {T.}~\bibnamefont {Monz}}, \bibinfo {author}
  {\bibfnamefont {P.}~\bibnamefont {Schindler}}, \bibinfo {author}
  {\bibfnamefont {B.}~\bibnamefont {Vogell}}, \bibinfo {author} {\bibfnamefont
  {M.}~\bibnamefont {Dalmonte}}, \bibinfo {author} {\bibfnamefont
  {P.}~\bibnamefont {Hauke}}, \bibinfo {author} {\bibfnamefont
  {R.}~\bibnamefont {Blatt}}, \ and\ \bibinfo {author} {\bibfnamefont
  {P.}~\bibnamefont {Zoller}},\ }\href {\doibase 10.1088/1367-2630/aa89ab}
  {\bibfield  {journal} {\bibinfo  {journal} {New J. Phys.}\ }\textbf {\bibinfo
  {volume} {19}},\ \bibinfo {pages} {103020} (\bibinfo {year} {2017})},\
  \Eprint {http://arxiv.org/abs/1612.08653} {arXiv:1612.08653 [quant-ph]}
  \BibitemShut {NoStop}%
\bibitem [{\citenamefont {Gustafson}\ \emph {et~al.}(2019)\citenamefont
  {Gustafson}, \citenamefont {Meurice},\ and\ \citenamefont
  {Unmuth-Yockey}}]{Gustafson:2019mpk}%
  \BibitemOpen
  \bibfield  {author} {\bibinfo {author} {\bibfnamefont {E.}~\bibnamefont
  {Gustafson}}, \bibinfo {author} {\bibfnamefont {Y.}~\bibnamefont {Meurice}},
  \ and\ \bibinfo {author} {\bibfnamefont {J.}~\bibnamefont {Unmuth-Yockey}},\
  }\href@noop {} {\  (\bibinfo {year} {2019})},\ \Eprint
  {http://arxiv.org/abs/1901.05944} {arXiv:1901.05944 [hep-lat]} \BibitemShut
  {NoStop}%
\bibitem [{\citenamefont {Zohar}\ \emph {et~al.}(2012)\citenamefont {Zohar},
  \citenamefont {Cirac},\ and\ \citenamefont {Reznik}}]{Zohar:2012ay}%
  \BibitemOpen
  \bibfield  {author} {\bibinfo {author} {\bibfnamefont {E.}~\bibnamefont
  {Zohar}}, \bibinfo {author} {\bibfnamefont {J.~I.}\ \bibnamefont {Cirac}}, \
  and\ \bibinfo {author} {\bibfnamefont {B.}~\bibnamefont {Reznik}},\ }\href
  {\doibase 10.1103/PhysRevLett.109.125302} {\bibfield  {journal} {\bibinfo
  {journal} {Phys. Rev. Lett.}\ }\textbf {\bibinfo {volume} {109}},\ \bibinfo
  {pages} {125302} (\bibinfo {year} {2012})},\ \Eprint
  {http://arxiv.org/abs/1204.6574} {arXiv:1204.6574 [quant-ph]} \BibitemShut
  {NoStop}%
\bibitem [{\citenamefont {Byrnes}\ and\ \citenamefont
  {Yamamoto}(2006)}]{Byrnes:2005qx}%
  \BibitemOpen
  \bibfield  {author} {\bibinfo {author} {\bibfnamefont {T.}~\bibnamefont
  {Byrnes}}\ and\ \bibinfo {author} {\bibfnamefont {Y.}~\bibnamefont
  {Yamamoto}},\ }\href {\doibase 10.1103/PhysRevA.73.022328} {\bibfield
  {journal} {\bibinfo  {journal} {Phys. Rev.}\ }\textbf {\bibinfo {volume}
  {A73}},\ \bibinfo {pages} {022328} (\bibinfo {year} {2006})},\ \Eprint
  {http://arxiv.org/abs/quant-ph/0510027} {arXiv:quant-ph/0510027 [quant-ph]}
  \BibitemShut {NoStop}%
\bibitem [{\citenamefont {Banerjee}\ \emph {et~al.}(2013)\citenamefont
  {Banerjee}, \citenamefont {Bögli}, \citenamefont {Dalmonte}, \citenamefont
  {Rico}, \citenamefont {Stebler}, \citenamefont {Wiese},\ and\ \citenamefont
  {Zoller}}]{Banerjee:2012xg}%
  \BibitemOpen
  \bibfield  {author} {\bibinfo {author} {\bibfnamefont {D.}~\bibnamefont
  {Banerjee}}, \bibinfo {author} {\bibfnamefont {M.}~\bibnamefont {Bögli}},
  \bibinfo {author} {\bibfnamefont {M.}~\bibnamefont {Dalmonte}}, \bibinfo
  {author} {\bibfnamefont {E.}~\bibnamefont {Rico}}, \bibinfo {author}
  {\bibfnamefont {P.}~\bibnamefont {Stebler}}, \bibinfo {author} {\bibfnamefont
  {U.~J.}\ \bibnamefont {Wiese}}, \ and\ \bibinfo {author} {\bibfnamefont
  {P.}~\bibnamefont {Zoller}},\ }\href {\doibase
  10.1103/PhysRevLett.110.125303} {\bibfield  {journal} {\bibinfo  {journal}
  {Phys. Rev. Lett.}\ }\textbf {\bibinfo {volume} {110}},\ \bibinfo {pages}
  {125303} (\bibinfo {year} {2013})},\ \Eprint {http://arxiv.org/abs/1211.2242}
  {arXiv:1211.2242 [cond-mat.quant-gas]} \BibitemShut {NoStop}%
\bibitem [{\citenamefont {Wiese}(2013)}]{Wiese:2013uua}%
  \BibitemOpen
  \bibfield  {author} {\bibinfo {author} {\bibfnamefont {U.-J.}\ \bibnamefont
  {Wiese}},\ }\href {\doibase 10.1002/andp.201300104} {\bibfield  {journal}
  {\bibinfo  {journal} {Annalen Phys.}\ }\textbf {\bibinfo {volume} {525}},\
  \bibinfo {pages} {777} (\bibinfo {year} {2013})},\ \Eprint
  {http://arxiv.org/abs/1305.1602} {arXiv:1305.1602 [quant-ph]} \BibitemShut
  {NoStop}%
\bibitem [{\citenamefont {Zohar}\ \emph {et~al.}(2013)\citenamefont {Zohar},
  \citenamefont {Cirac},\ and\ \citenamefont {Reznik}}]{Zohar:2012xf}%
  \BibitemOpen
  \bibfield  {author} {\bibinfo {author} {\bibfnamefont {E.}~\bibnamefont
  {Zohar}}, \bibinfo {author} {\bibfnamefont {J.~I.}\ \bibnamefont {Cirac}}, \
  and\ \bibinfo {author} {\bibfnamefont {B.}~\bibnamefont {Reznik}},\ }\href
  {\doibase 10.1103/PhysRevLett.110.125304} {\bibfield  {journal} {\bibinfo
  {journal} {Phys. Rev. Lett.}\ }\textbf {\bibinfo {volume} {110}},\ \bibinfo
  {pages} {125304} (\bibinfo {year} {2013})},\ \Eprint
  {http://arxiv.org/abs/1211.2241} {arXiv:1211.2241 [quant-ph]} \BibitemShut
  {NoStop}%
\bibitem [{\citenamefont {Tagliacozzo}\ \emph {et~al.}(2013)\citenamefont
  {Tagliacozzo}, \citenamefont {Celi}, \citenamefont {Orland},\ and\
  \citenamefont {Lewenstein}}]{Tagliacozzo:2012df}%
  \BibitemOpen
  \bibfield  {author} {\bibinfo {author} {\bibfnamefont {L.}~\bibnamefont
  {Tagliacozzo}}, \bibinfo {author} {\bibfnamefont {A.}~\bibnamefont {Celi}},
  \bibinfo {author} {\bibfnamefont {P.}~\bibnamefont {Orland}}, \ and\ \bibinfo
  {author} {\bibfnamefont {M.}~\bibnamefont {Lewenstein}},\ }\href {\doibase
  10.1038/ncomms3615} {\bibfield  {journal} {\bibinfo  {journal} {Nature
  Commun.}\ }\textbf {\bibinfo {volume} {4}},\ \bibinfo {pages} {2615}
  (\bibinfo {year} {2013})},\ \Eprint {http://arxiv.org/abs/1211.2704}
  {arXiv:1211.2704 [cond-mat.quant-gas]} \BibitemShut {NoStop}%
\bibitem [{\citenamefont {Zohar}\ \emph {et~al.}(2016)\citenamefont {Zohar},
  \citenamefont {Cirac},\ and\ \citenamefont {Reznik}}]{Zohar:2015hwa}%
  \BibitemOpen
  \bibfield  {author} {\bibinfo {author} {\bibfnamefont {E.}~\bibnamefont
  {Zohar}}, \bibinfo {author} {\bibfnamefont {J.~I.}\ \bibnamefont {Cirac}}, \
  and\ \bibinfo {author} {\bibfnamefont {B.}~\bibnamefont {Reznik}},\ }\href
  {\doibase 10.1088/0034-4885/79/1/014401} {\bibfield  {journal} {\bibinfo
  {journal} {Rept. Prog. Phys.}\ }\textbf {\bibinfo {volume} {79}},\ \bibinfo
  {pages} {014401} (\bibinfo {year} {2016})},\ \Eprint
  {http://arxiv.org/abs/1503.02312} {arXiv:1503.02312 [quant-ph]} \BibitemShut
  {NoStop}%
\bibitem [{\citenamefont {Zohar}\ \emph {et~al.}(2017)\citenamefont {Zohar},
  \citenamefont {Farace}, \citenamefont {Reznik},\ and\ \citenamefont
  {Cirac}}]{Zohar:2016iic}%
  \BibitemOpen
  \bibfield  {author} {\bibinfo {author} {\bibfnamefont {E.}~\bibnamefont
  {Zohar}}, \bibinfo {author} {\bibfnamefont {A.}~\bibnamefont {Farace}},
  \bibinfo {author} {\bibfnamefont {B.}~\bibnamefont {Reznik}}, \ and\ \bibinfo
  {author} {\bibfnamefont {J.~I.}\ \bibnamefont {Cirac}},\ }\href {\doibase
  10.1103/PhysRevA.95.023604} {\bibfield  {journal} {\bibinfo  {journal} {Phys.
  Rev.}\ }\textbf {\bibinfo {volume} {A95}},\ \bibinfo {pages} {023604}
  (\bibinfo {year} {2017})},\ \Eprint {http://arxiv.org/abs/1607.08121}
  {arXiv:1607.08121 [quant-ph]} \BibitemShut {NoStop}%
\bibitem [{\citenamefont {Bender}\ \emph {et~al.}(2018)\citenamefont {Bender},
  \citenamefont {Zohar}, \citenamefont {Farace},\ and\ \citenamefont
  {Cirac}}]{Bender:2018rdp}%
  \BibitemOpen
  \bibfield  {author} {\bibinfo {author} {\bibfnamefont {J.}~\bibnamefont
  {Bender}}, \bibinfo {author} {\bibfnamefont {E.}~\bibnamefont {Zohar}},
  \bibinfo {author} {\bibfnamefont {A.}~\bibnamefont {Farace}}, \ and\ \bibinfo
  {author} {\bibfnamefont {J.~I.}\ \bibnamefont {Cirac}},\ }\href {\doibase
  10.1088/1367-2630/aadb71} {\bibfield  {journal} {\bibinfo  {journal} {New J.
  Phys.}\ }\textbf {\bibinfo {volume} {20}},\ \bibinfo {pages} {093001}
  (\bibinfo {year} {2018})},\ \Eprint {http://arxiv.org/abs/1804.02082}
  {arXiv:1804.02082 [quant-ph]} \BibitemShut {NoStop}%
\bibitem [{\citenamefont {Stryker}(2018)}]{Stryker:2018efp}%
  \BibitemOpen
  \bibfield  {author} {\bibinfo {author} {\bibfnamefont {J.~R.}\ \bibnamefont
  {Stryker}},\ }\href@noop {} {\  (\bibinfo {year} {2018})},\ \Eprint
  {http://arxiv.org/abs/1812.01617} {arXiv:1812.01617 [quant-ph]} \BibitemShut
  {NoStop}%
\bibitem [{\citenamefont {Luscher}(1977)}]{Luscher:1976ms}%
  \BibitemOpen
  \bibfield  {author} {\bibinfo {author} {\bibfnamefont {M.}~\bibnamefont
  {Luscher}},\ }\href {\doibase 10.1007/BF01614090} {\bibfield  {journal}
  {\bibinfo  {journal} {Commun. Math. Phys.}\ }\textbf {\bibinfo {volume}
  {54}},\ \bibinfo {pages} {283} (\bibinfo {year} {1977})}\BibitemShut
  {NoStop}%
\bibitem [{\citenamefont {Mitrjushkin}(2002)}]{Mitrjushkin:2002nk}%
  \BibitemOpen
  \bibfield  {author} {\bibinfo {author} {\bibfnamefont {V.~K.}\ \bibnamefont
  {Mitrjushkin}},\ }\href@noop {} {\  (\bibinfo {year} {2002})},\ \Eprint
  {http://arxiv.org/abs/0206024} {arXiv:0206024 [hep-lat]} \BibitemShut
  {NoStop}%
\bibitem [{\citenamefont {Creutz}(2000)}]{Creutz:1999zy}%
  \BibitemOpen
  \bibfield  {author} {\bibinfo {author} {\bibfnamefont {M.}~\bibnamefont
  {Creutz}},\ }\href {\doibase 10.1023/A:1003630124933} {\bibfield  {journal}
  {\bibinfo  {journal} {Found. Phys.}\ }\textbf {\bibinfo {volume} {30}},\
  \bibinfo {pages} {487} (\bibinfo {year} {2000})},\ \Eprint
  {http://arxiv.org/abs/9905024} {arXiv:9905024 [hep-lat]} \BibitemShut
  {NoStop}%
\bibitem [{\citenamefont {Caracciolo}\ and\ \citenamefont
  {Palumbo}(2013)}]{Caracciolo:2012qw}%
  \BibitemOpen
  \bibfield  {author} {\bibinfo {author} {\bibfnamefont {S.}~\bibnamefont
  {Caracciolo}}\ and\ \bibinfo {author} {\bibfnamefont {F.}~\bibnamefont
  {Palumbo}},\ }\href {\doibase 10.1103/PhysRevD.87.014507} {\bibfield
  {journal} {\bibinfo  {journal} {Phys. Rev.}\ }\textbf {\bibinfo {volume}
  {D87}},\ \bibinfo {pages} {014507} (\bibinfo {year} {2013})},\ \Eprint
  {http://arxiv.org/abs/1210.1786} {arXiv:1210.1786 [hep-lat]} \BibitemShut
  {NoStop}%
\bibitem [{\citenamefont {Haegeman}\ \emph {et~al.}(2013)\citenamefont
  {Haegeman}, \citenamefont {Osborne}, \citenamefont {Verschelde},\ and\
  \citenamefont {Verstraete}}]{Haegeman:2011uy}%
  \BibitemOpen
  \bibfield  {author} {\bibinfo {author} {\bibfnamefont {J.}~\bibnamefont
  {Haegeman}}, \bibinfo {author} {\bibfnamefont {T.~J.}\ \bibnamefont
  {Osborne}}, \bibinfo {author} {\bibfnamefont {H.}~\bibnamefont {Verschelde}},
  \ and\ \bibinfo {author} {\bibfnamefont {F.}~\bibnamefont {Verstraete}},\
  }\href {\doibase 10.1103/PhysRevLett.110.100402} {\bibfield  {journal}
  {\bibinfo  {journal} {Phys. Rev. Lett.}\ }\textbf {\bibinfo {volume} {110}},\
  \bibinfo {pages} {100402} (\bibinfo {year} {2013})},\ \Eprint
  {http://arxiv.org/abs/1102.5524} {arXiv:1102.5524 [hep-th]} \BibitemShut
  {NoStop}%
\bibitem [{\citenamefont {Verstraete}\ and\ \citenamefont
  {Cirac}(2010)}]{Verstraete:2010ft}%
  \BibitemOpen
  \bibfield  {author} {\bibinfo {author} {\bibfnamefont {F.}~\bibnamefont
  {Verstraete}}\ and\ \bibinfo {author} {\bibfnamefont {J.~I.}\ \bibnamefont
  {Cirac}},\ }\href {\doibase 10.1103/PhysRevLett.104.190405} {\bibfield
  {journal} {\bibinfo  {journal} {Phys. Rev. Lett.}\ }\textbf {\bibinfo
  {volume} {104}},\ \bibinfo {pages} {190405} (\bibinfo {year} {2010})},\
  \Eprint {http://arxiv.org/abs/1002.1824} {arXiv:1002.1824 [cond-mat.str-el]}
  \BibitemShut {NoStop}%
\bibitem [{\citenamefont {Rossi}\ and\ \citenamefont
  {Testa}(2018)}]{Rossi:2018zkn}%
  \BibitemOpen
  \bibfield  {author} {\bibinfo {author} {\bibfnamefont {G.}~\bibnamefont
  {Rossi}}\ and\ \bibinfo {author} {\bibfnamefont {M.}~\bibnamefont {Testa}},\
  }\href {\doibase 10.1103/PhysRevD.98.054028} {\bibfield  {journal} {\bibinfo
  {journal} {Phys. Rev.}\ }\textbf {\bibinfo {volume} {D98}},\ \bibinfo {pages}
  {054028} (\bibinfo {year} {2018})},\ \Eprint
  {http://arxiv.org/abs/1806.00808} {arXiv:1806.00808 [hep-lat]} \BibitemShut
  {NoStop}%
\bibitem [{\citenamefont {Briceño}\ \emph {et~al.}(2017)\citenamefont
  {Briceño}, \citenamefont {Hansen},\ and\ \citenamefont
  {Monahan}}]{Briceno:2017cpo}%
  \BibitemOpen
  \bibfield  {author} {\bibinfo {author} {\bibfnamefont {R.~A.}\ \bibnamefont
  {Briceño}}, \bibinfo {author} {\bibfnamefont {M.~T.}\ \bibnamefont
  {Hansen}}, \ and\ \bibinfo {author} {\bibfnamefont {C.~J.}\ \bibnamefont
  {Monahan}},\ }\href {\doibase 10.1103/PhysRevD.96.014502} {\bibfield
  {journal} {\bibinfo  {journal} {Phys. Rev.}\ }\textbf {\bibinfo {volume}
  {D96}},\ \bibinfo {pages} {014502} (\bibinfo {year} {2017})},\ \Eprint
  {http://arxiv.org/abs/1703.06072} {arXiv:1703.06072 [hep-lat]} \BibitemShut
  {NoStop}%
\bibitem [{\citenamefont {Cotler}\ \emph
  {et~al.}(2018{\natexlab{a}})\citenamefont {Cotler}, \citenamefont
  {Mohammadi~Mozaffar}, \citenamefont {Mollabashi},\ and\ \citenamefont
  {Naseh}}]{Cotler:2018ehb}%
  \BibitemOpen
  \bibfield  {author} {\bibinfo {author} {\bibfnamefont {J.}~\bibnamefont
  {Cotler}}, \bibinfo {author} {\bibfnamefont {M.~R.}\ \bibnamefont
  {Mohammadi~Mozaffar}}, \bibinfo {author} {\bibfnamefont {A.}~\bibnamefont
  {Mollabashi}}, \ and\ \bibinfo {author} {\bibfnamefont {A.}~\bibnamefont
  {Naseh}},\ }\href@noop {} {\  (\bibinfo {year} {2018}{\natexlab{a}})},\
  \Eprint {http://arxiv.org/abs/1806.02835} {arXiv:1806.02835 [hep-th]}
  \BibitemShut {NoStop}%
\bibitem [{\citenamefont {Cotler}\ \emph
  {et~al.}(2018{\natexlab{b}})\citenamefont {Cotler}, \citenamefont
  {Mohammadi~Mozaffar}, \citenamefont {Mollabashi},\ and\ \citenamefont
  {Naseh}}]{Cotler:2018ufx}%
  \BibitemOpen
  \bibfield  {author} {\bibinfo {author} {\bibfnamefont {J.}~\bibnamefont
  {Cotler}}, \bibinfo {author} {\bibfnamefont {M.~R.}\ \bibnamefont
  {Mohammadi~Mozaffar}}, \bibinfo {author} {\bibfnamefont {A.}~\bibnamefont
  {Mollabashi}}, \ and\ \bibinfo {author} {\bibfnamefont {A.}~\bibnamefont
  {Naseh}},\ }\href@noop {} {\  (\bibinfo {year} {2018}{\natexlab{b}})},\
  \Eprint {http://arxiv.org/abs/1806.02831} {arXiv:1806.02831 [hep-th]}
  \BibitemShut {NoStop}%
\bibitem [{\citenamefont {Coppersmith}(2002)}]{coppersmith2002approximate}%
  \BibitemOpen
  \bibfield  {author} {\bibinfo {author} {\bibfnamefont {D.}~\bibnamefont
  {Coppersmith}},\ }\href@noop {} {\bibfield  {journal} {\bibinfo  {journal}
  {arXiv:0201067}\ } (\bibinfo {year} {2002})}\BibitemShut {NoStop}%
\bibitem [{\citenamefont {Ekert}\ and\ \citenamefont
  {Jozsa}(1996)}]{RevModPhys.68.733}%
  \BibitemOpen
  \bibfield  {author} {\bibinfo {author} {\bibfnamefont {A.}~\bibnamefont
  {Ekert}}\ and\ \bibinfo {author} {\bibfnamefont {R.}~\bibnamefont {Jozsa}},\
  }\href {\doibase 10.1103/RevModPhys.68.733} {\bibfield  {journal} {\bibinfo
  {journal} {Rev. Mod. Phys.}\ }\textbf {\bibinfo {volume} {68}},\ \bibinfo
  {pages} {733} (\bibinfo {year} {1996})}\BibitemShut {NoStop}%
\bibitem [{\citenamefont {Jozsa}(1998)}]{jozsa1998quantum}%
  \BibitemOpen
  \bibfield  {author} {\bibinfo {author} {\bibfnamefont {R.}~\bibnamefont
  {Jozsa}},\ }in\ \href@noop {} {\emph {\bibinfo {booktitle} {Proceedings of
  the Royal Society of London A: Mathematical, Physical and Engineering
  Sciences}}},\ Vol.\ \bibinfo {volume} {454}\ (\bibinfo {organization} {The
  Royal Society},\ \bibinfo {year} {1998})\ pp.\ \bibinfo {pages}
  {323--337}\BibitemShut {NoStop}%
\bibitem [{\citenamefont {Zohar}(2018)}]{Zohar:2018nvl}%
  \BibitemOpen
  \bibfield  {author} {\bibinfo {author} {\bibfnamefont {E.}~\bibnamefont
  {Zohar}},\ }in\ \href@noop {} {\emph {\bibinfo {booktitle} {{Tensor Network
  and entanglement Florence, Italy, June 18-22, 2018}}}}\ (\bibinfo {year}
  {2018})\ \Eprint {http://arxiv.org/abs/1807.01294} {arXiv:1807.01294
  [quant-ph]} \BibitemShut {NoStop}%
\bibitem [{\citenamefont {Kogut}\ and\ \citenamefont
  {Susskind}(1975)}]{PhysRevD.11.395}%
  \BibitemOpen
  \bibfield  {author} {\bibinfo {author} {\bibfnamefont {J.}~\bibnamefont
  {Kogut}}\ and\ \bibinfo {author} {\bibfnamefont {L.}~\bibnamefont
  {Susskind}},\ }\href {\doibase 10.1103/PhysRevD.11.395} {\bibfield  {journal}
  {\bibinfo  {journal} {Phys. Rev. D}\ }\textbf {\bibinfo {volume} {11}},\
  \bibinfo {pages} {395} (\bibinfo {year} {1975})}\BibitemShut {NoStop}%
\bibitem [{\citenamefont {Harlow}\ and\ \citenamefont
  {Ooguri}(2018)}]{Harlow:2018tng}%
  \BibitemOpen
  \bibfield  {author} {\bibinfo {author} {\bibfnamefont {D.}~\bibnamefont
  {Harlow}}\ and\ \bibinfo {author} {\bibfnamefont {H.}~\bibnamefont
  {Ooguri}},\ }\href@noop {} {\  (\bibinfo {year} {2018})},\ \Eprint
  {http://arxiv.org/abs/1810.05338} {arXiv:1810.05338 [hep-th]} \BibitemShut
  {NoStop}%
\bibitem [{\citenamefont {Creutz}(1977)}]{Creutz:1976ch}%
  \BibitemOpen
  \bibfield  {author} {\bibinfo {author} {\bibfnamefont {M.}~\bibnamefont
  {Creutz}},\ }\href {\doibase 10.1103/PhysRevD.15.1128} {\bibfield  {journal}
  {\bibinfo  {journal} {Phys. Rev.}\ }\textbf {\bibinfo {volume} {D15}},\
  \bibinfo {pages} {1128} (\bibinfo {year} {1977})}\BibitemShut {NoStop}%
\bibitem [{\citenamefont {Ortiz}\ \emph {et~al.}(2001)\citenamefont {Ortiz},
  \citenamefont {Gubernatis}, \citenamefont {Knill},\ and\ \citenamefont
  {Laflamme}}]{Ortiz:2000gc}%
  \BibitemOpen
  \bibfield  {author} {\bibinfo {author} {\bibfnamefont {G.}~\bibnamefont
  {Ortiz}}, \bibinfo {author} {\bibfnamefont {J.~E.}\ \bibnamefont
  {Gubernatis}}, \bibinfo {author} {\bibfnamefont {E.}~\bibnamefont {Knill}}, \
  and\ \bibinfo {author} {\bibfnamefont {R.}~\bibnamefont {Laflamme}},\ }\href
  {\doibase 10.1103/PhysRevA.64.022319} {\bibfield  {journal} {\bibinfo
  {journal} {Phys. Rev.}\ }\textbf {\bibinfo {volume} {A64}},\ \bibinfo {pages}
  {022319} (\bibinfo {year} {2001})},\ \Eprint
  {http://arxiv.org/abs/cond-mat/0012334} {arXiv:cond-mat/0012334 [cond-mat]}
  \BibitemShut {NoStop}%
\bibitem [{\citenamefont {Santos}(2017)}]{santos2017ibm}%
  \BibitemOpen
  \bibfield  {author} {\bibinfo {author} {\bibfnamefont {A.~C.}\ \bibnamefont
  {Santos}},\ }\href@noop {} {\bibfield  {journal} {\bibinfo  {journal}
  {Revista Brasileira de Ensino de F{\'\i}sica}\ }\textbf {\bibinfo {volume}
  {39}} (\bibinfo {year} {2017})},\ \Eprint {http://arxiv.org/abs/1610.06980}
  {arXiv:1610.06980 [quant-ph]} \BibitemShut {NoStop}%
\bibitem [{\citenamefont {Cross}\ \emph {et~al.}(2017)\citenamefont {Cross},
  \citenamefont {Bishop}, \citenamefont {Smolin},\ and\ \citenamefont
  {Gambetta}}]{cross2017open}%
  \BibitemOpen
  \bibfield  {author} {\bibinfo {author} {\bibfnamefont {A.~W.}\ \bibnamefont
  {Cross}}, \bibinfo {author} {\bibfnamefont {L.~S.}\ \bibnamefont {Bishop}},
  \bibinfo {author} {\bibfnamefont {J.~A.}\ \bibnamefont {Smolin}}, \ and\
  \bibinfo {author} {\bibfnamefont {J.~M.}\ \bibnamefont {Gambetta}},\
  }\href@noop {} {\bibfield  {journal} {\bibinfo  {journal} {arXiv:1707.03429}\
  } (\bibinfo {year} {2017})}\BibitemShut {NoStop}%
\bibitem [{\citenamefont {Nam}\ \emph {et~al.}(2019)\citenamefont {Nam},
  \citenamefont {Chen}, \citenamefont {Pisenti}, \citenamefont {Wright},
  \citenamefont {Delaney}, \citenamefont {Maslov}, \citenamefont {Brown},
  \citenamefont {Allen}, \citenamefont {Amini}, \citenamefont {Apisdorf} \emph
  {et~al.}}]{nam2019ground}%
  \BibitemOpen
  \bibfield  {author} {\bibinfo {author} {\bibfnamefont {Y.}~\bibnamefont
  {Nam}}, \bibinfo {author} {\bibfnamefont {J.-S.}\ \bibnamefont {Chen}},
  \bibinfo {author} {\bibfnamefont {N.~C.}\ \bibnamefont {Pisenti}}, \bibinfo
  {author} {\bibfnamefont {K.}~\bibnamefont {Wright}}, \bibinfo {author}
  {\bibfnamefont {C.}~\bibnamefont {Delaney}}, \bibinfo {author} {\bibfnamefont
  {D.}~\bibnamefont {Maslov}}, \bibinfo {author} {\bibfnamefont {K.~R.}\
  \bibnamefont {Brown}}, \bibinfo {author} {\bibfnamefont {S.}~\bibnamefont
  {Allen}}, \bibinfo {author} {\bibfnamefont {J.~M.}\ \bibnamefont {Amini}},
  \bibinfo {author} {\bibfnamefont {J.}~\bibnamefont {Apisdorf}},  \emph
  {et~al.},\ }\href@noop {} {\bibfield  {journal} {\bibinfo  {journal}
  {arXiv:1902.10171}\ } (\bibinfo {year} {2019})}\BibitemShut {NoStop}%
\bibitem [{\citenamefont {{Moore}}\ \emph {et~al.}(2003)\citenamefont
  {{Moore}}, \citenamefont {{Rockmore}},\ and\ \citenamefont
  {{Russell}}}]{2003quant.ph..4064M}%
  \BibitemOpen
  \bibfield  {author} {\bibinfo {author} {\bibfnamefont {C.}~\bibnamefont
  {{Moore}}}, \bibinfo {author} {\bibfnamefont {D.}~\bibnamefont {{Rockmore}}},
  \ and\ \bibinfo {author} {\bibfnamefont {A.}~\bibnamefont {{Russell}}},\
  }\href@noop {} {\  (\bibinfo {year} {2003})},\ \Eprint
  {http://arxiv.org/abs/0304064} {arXiv:0304064 [quant-ph]} \BibitemShut
  {NoStop}%
\bibitem [{\citenamefont {{Pueschel}}\ \emph {et~al.}(1998)\citenamefont
  {{Pueschel}}, \citenamefont {{Roetteler}},\ and\ \citenamefont
  {{Beth}}}]{1998quant.ph..7064P}%
  \BibitemOpen
  \bibfield  {author} {\bibinfo {author} {\bibfnamefont {M.}~\bibnamefont
  {{Pueschel}}}, \bibinfo {author} {\bibfnamefont {M.}~\bibnamefont
  {{Roetteler}}}, \ and\ \bibinfo {author} {\bibfnamefont {T.}~\bibnamefont
  {{Beth}}},\ }\href@noop {} {\ ,\ \bibinfo {eid} {quant-ph/9807064} (\bibinfo
  {year} {1998})},\ \Eprint {http://arxiv.org/abs/9807064} {arXiv:9807064
  [quant-ph]} \BibitemShut {NoStop}%
\bibitem [{\citenamefont {Beals}(1997)}]{beals1997quantum}%
  \BibitemOpen
  \bibfield  {author} {\bibinfo {author} {\bibfnamefont {R.}~\bibnamefont
  {Beals}},\ }in\ \href@noop {} {\emph {\bibinfo {booktitle} {Proceedings of
  the twenty-ninth annual ACM symposium on Theory of computing}}}\ (\bibinfo
  {organization} {Citeseer},\ \bibinfo {year} {1997})\ pp.\ \bibinfo {pages}
  {48--53}\BibitemShut {NoStop}%
\end{thebibliography}%
\end{document}